\documentclass[twocolumn]{openjournal}

\usepackage{graphicx}
\usepackage{amsmath}
\usepackage{gensymb}
\usepackage[super]{nth}
\usepackage{xspace}
\usepackage[dvipsnames]{xcolor}
\usepackage{bm}

\usepackage{hyperref}
\hypersetup{colorlinks=true, allcolors=Blue}

\newcommand{\lenstronomy}{\texttt{lenstronomy}\xspace}
\newcommand{\dolphin}{\texttt{dolphin}\xspace}
\newcommand{\amplification}{\bm{\mathcal{A}}}

\newcommand\e[1]{_{\mathrm{#1}}}
\newcommand\h[1]{^{\mathrm{#1}}}

\newcommand{\dd}{\mathrm{d}}

\newcommand{\ddf}[3][]{\frac{\dd^{#1} #2}{\dd {#3}^{#1}}}

\begin{document}
\title{Line-of-sight shear in \emph{SLACS} strong lenses I:\\shear and mass model parametrisations}

\author{Natalie B. Hogg$^{1, 2, 3, \ast}$,  Daniel P. Johnson$^4$, Anowar J. Shajib$^{5, 6, 7}$, Julien Larena$^4$}
\thanks{$^\ast$natalie.hogg@ast.cam.ac.uk}

\affiliation{$^1$ Institute of Astronomy,  University of Cambridge, Madingley Road, Cambridge, CB3 0HA, UK}
\affiliation{$^2$ Kavli Institute for Cosmology, University of Cambridge, Cambridge, UK}
\affiliation{$^3$ Clare Hall, Herschel Road, Cambridge, CB3 9AL, UK}
\affiliation{$^4$Laboratoire Univers et Particules de Montpellier, CNRS \& Université de Montpellier,\\Parvis Alexander Grothendieck, Montpellier, France, 34090}
\affiliation{$^5$Department of Astronomy \& Astrophysics, University of Chicago, Chicago, IL 60637, USA}
\affiliation{$^6$Kavli Institute for Cosmological Physics, University of Chicago, Chicago, IL 60637, USA}
\affiliation{$^7$Center for Astronomy, Space Science and Astrophysics, Independent University, Bangladesh, Dhaka 1229, Bangladesh}

\begin{abstract}
Inhomogeneities along the line of sight in strong gravitational lensing distort the images produced, in an effect called shear. If measurable, this shear may provide independent constraints on cosmological parameters, complementary to traditional cosmic shear. We model 23 strong gravitational lenses from the Sloan Lens ACS (SLACS) catalogue with the aim of measuring the line-of-sight (LOS) shear for the first time. We use the `minimal model' for the LOS shear, which has been shown to be theoretically safe from degeneracies with lens model parameters, a finding which has been confirmed using mock data. We use the \texttt{dolphin} automated modelling pipeline, which uses the \texttt{lenstronomy} software as a modelling engine, to model our selected lenses. Across the 23 lenses, we measure the LOS shear with a mean magnitude of $0.056 \pm 0.013$. Neglecting the post-Born correction to the potential of the main deflector due to foreground shear leads to a propagation of degeneracies in the LOS shear measurement with other lens model parameters, and the inclusion of an octupole moment in the lens mass profile does not lead to shear measurements that are in better agreement with the expectations from weak lensing.
\end{abstract}


\section{Introduction}
Gravitational lensing provides a unique window into the cosmology of our Universe on a wide range of scales. The strong lensing regime, in which a single massive deflector lenses a single source, has been used to measure the expansion rate of the Universe at $z=0$, $H_0$ \citep{Refsdal1964b, Wong:2019kwg, Birrer:2022chj, TDCOSMO:2025dmr} via observations of the difference in arrival times of images of variable sources. Strong lensing has been used to probe properties of the lens galaxies, such as their density profiles \citep{Treu2006,Shajib2021, Shajib2024}, and, by studying lens galaxies at different redshifts, their evolution over time \citep{Chae2010,Bezanson2011,Sahu:2024ayh, Sheu:2024sjd}. It has also been used to search for dark matter sub-haloes both within the main deflector halo and along the line of sight between the observer and lens \citep{Vegetti_2009, Vegetti_2012, Vegetti_2014, Hezaveh:2016ltk, Nightingale2024b}.

The weak lensing regime probes larger scales. Unlike strong lensing, multiple images of the same source are not produced; rather, small distortions are induced on the single image of each source, artificially squashing and aligning them. The leading order shape distortion is shear, and its effect cannot be disentangled from the intrinsic shape of individual galaxies or their intrinsic alignment. The cosmic shear signal is detected by correlating the shapes of millions of galaxies in a given galaxy survey. Once extracted, cosmic shear can be used to place constraints on cosmological parameters like the matter density $\Omega_{\rm m}$ and the clustering parameter $\sigma_8$. For a recent review of weak gravitational lensing, we refer the reader to \cite{Prat:2025ucy}.

By virtue of their existence in our inhomogeneous Universe, strong lensing images will also experience weak lensing distortions, with Einstein rings in particular supplying the notion of a `standardisable shape' from which shear may be robustly measured\footnote{In practice, shear can be recovered even when the image is not a complete Einstein ring, though the precision of the measurement may be degraded \citep{Hogg:2022ycw}.}, providing an alternative and independent probe of cosmology to the standard weak lensing shear. 

There is a twofold benefit to using strong lensing shear data in cosmological analyses: firstly, simply by virtue of enlarging the data vector, there will be an automatic increase in statistical power when constraining cosmological parameters with lensing data; and secondly, whilst systematic uncertainties will undoubtedly be present in such constraints, they will nonetheless be complementary, rather than additive, to the systematics present in standard weak lensing data, such as galaxy bias and intrinsic alignments. The most obvious systematics, shear and lens mass model choice and their subsequent effect on shear measurements, are the focus of this work.

The idea of using strong lensing images to measure shear for cosmology was first investigated by \cite{Birrer:2016xku}, who modelled the lens COSMOS 0038+4133 with a shear model that consisted of four parameters: two components to describe the foreground shear as projected onto the lens plane, and two components to describe the background shear as projected onto the source plane. Whilst the shear parameters were shown to be constrained with high precision, this analysis revealed strong degeneracies between the shear parameters and the components of the ellipticity of the lens mass. This implies that, with this model, the constraints on the shear cannot be disentangled from the constraints on the lens mass, and the measurements cannot be safely used to obtain cosmological information.

\cite{Fleury:2021tke} developed the minimal line-of-sight shear model to address this problem. Starting with three separate shears, the model exploits a source-position transformation to build a shear parameter, the `line-of-sight (LOS) shear', which is mathematically free from degeneracy with ellipticity in the lens mass. The model fully describes the problem with the fewest possible parameters, and is therefore called the \textit{minimal} LOS shear model -- we will review the formalism behind this model in more detail below. Thanks to this freedom from degeneracies, it is the LOS shear in the minimal model that should be targeted for eventual cosmological inference. It was shown by \cite{Hogg:2022ycw} that the LOS shear is indeed free from degeneracies with lens model parameters using mock strong lensing images. It is therefore timely to attempt to measure this quantity in real strong lens imaging data. 

A further motivation of this work is to explore the effect of different shear and mass model parametrisations, in light of the findings of \cite{Etherington:2023yyh}, who showed that external shear magnitudes in some strong lenses are suspiciously large. They proposed that unmodelled complexity in the lens mass may be the reason for this. We therefore explore the role of foreground shear in the minimal LOS model, as well as octupolar distortions in the lens mass, in the sample of lenses studied in this work. 

We note that a number of previous works have also investigated shear systematics in various contexts; for example, \cite{Witt1997} and \cite{Gomer2018} studied quadruply imaged systems at the individual and population level, finding that image positions cannot be reproduced in the absence of significant external shear; \cite{Wong2011} studied the local environments of six quads to estimate external shear, finding discrepancies in the magnitude and direction compared to the shears obtained from lens modelling in three of the six cases. \cite{Gomer:2021gio} showed with mock images that using a purely elliptical mass model plus shear to model lens masses containing deviations from ellipticity would result in a bias on the Hubble parameter, $H_0$, of $\sim 10$\%. \cite{Barrera2021} emphasised that both parametrised and free-form mass models can provide reasonable fits to the quad WFI2033$-$4732, both favouring lopsidedness in the mass, but that the astrophysical origin of such a lopsidedness (internal to the lens, within the local environment, or on the line of sight) remains difficult to determine, referring to this as the `interpretation degeneracy'. Whilst these works mainly focused on quads in the context of determining $H_0$ from time delays, the conclusions should also apply to the lenses with multiply imaged extended sources that we consider in this work.

This paper is organised as follows: in \autoref{sec:theory} we review strong lensing theory and the minimal model; in \autoref{sec:methods} we discuss the data used in this work and the method used to analyse it; in \autoref{sec:results} we present and discuss our results, namely fitting our sample of strong lenses with two different shear parametrisations and two different mass model parametrisations, including an octupolar distortion; we conclude with \autoref{sec:conclusions}. 

\section{Strong lensing and the line-of-sight shear} \label{sec:theory}
In this section we review the strong lensing formalism and the derivation of the minimal model for the LOS shear. For full details, the reader is referred to \cite{Fleury:2021tke}. 

The path of a light ray through the Universe is perturbed by the presence of massive objects. In the simplest possible case, there is a single deflector that perturbs the ray, with the angular position of the observed image $\bm{\theta}$ related to that of the unobserved source $\bm{\beta}$ via the lens equation,
\begin{equation}
	\bm{\beta} = \bm{\theta} -  \bm{\alpha}(\bm{\theta}), \label{eq:single_plane_lens}
\end{equation}
where $\bm{\alpha}(\bm{\theta})$ is the displacement angle (also called the deflection angle), which depends on the gravitational potential of the deflector projected along the LOS into a single plane, i.e.
\begin{equation}
	\bm{\alpha}(\bm{\theta})
	= \ddf{\psi}{\bm{\theta}} \ ,
\end{equation}
where
\begin{equation}
	\psi(\bm{\theta})
	\equiv
	\frac{D\e{ds}}{D\e{od}D\e{os}} \, \hat{\psi}(D\e{od}\bm{\theta}) \ ,
\end{equation}
and $\hat{\psi}(\bm{x})$ is twice the projected gravitational potential produced by the surface density of the main lens, $\Sigma(\bm{x})$,
\begin{equation}
	\hat{\psi} (\bm{x})
	\equiv
	\frac{4G}{c^2} \int \dd^2 \bm{y} \; \Sigma(\bm{y}) \ln |\bm{x} - \bm{y}| \ ,
\end{equation}
where $\bm{x}$ is the vector describing the point where the light ray strikes the lens plane, $G$ is Newton's constant and $c$ the speed of light.

To good approximation, further inhomogeneities beyond the main deflector can be treated as tidal perturbations, which induce a shear and convergence on the strong lens image. The exceptions to this are, for example, double plane lensing \citep{Gavazzi:2008aq, Collett:2020lii, Dux:2024vvq} or systems for which nearby satellite galaxies are explicitly included in the model, e.g. \cite{Nightingale:2022bhh}. Non-tidal perturbations will also induce flexion on strong lensing images \citep{Goldberg:2004hh, Bacon:2005qr, Duboscq:2024asf}.

In the case of single plane lensing with tidal perturbations, \autoref{eq:single_plane_lens} becomes 
\begin{equation}
\bm{\beta} = \amplification\e{os} \bm{\theta} - \amplification\e{ds} \, \frac{ \dd \psi(\amplification\e{od} \bm{\theta})}{\dd \bm{\theta}},
\label{eq:lens_eqn_los}
\end{equation}
where the amplification matrices $\amplification\e{os}$, $\amplification\e{ds}$ and $\amplification\e{od}$ encapsulate the LOS perturbations between observer and source, main deflector and source, and observer and main deflector respectively. They can be parametrised as
\begin{align}
	\label{eq:decomposition_amplification_matrix}
	\amplification\e{ab} &= \bm{1} - \bm{\Gamma}\e{ab} \ ,
	\\
	\bm{\Gamma}\e{ab} &=
	\begin{bmatrix}
		\kappa\e{ab} + \rm{Re}\left(\gamma\e{ab}\right) & \rm{Im}\left(\gamma\e{ab}\right) - \omega\e{ab} \\ \rm{Im}\left(\gamma\e{ab}\right) + \omega\e{ab} & \kappa\e{ab}- \rm{Re}\left(\gamma\e{ab}\right)
	\end{bmatrix},
\end{align}
where $\mathrm{ab}\in\{\mathrm{os}, \mathrm{ds}, \mathrm{od}\}$ and $\kappa_{\rm ab}$ is the convergence, $\gamma_{\rm ab}$ the shear and $\omega_{\rm ab}$ the rotation of an image as seen at position `a' with respect to its source at `b', produced by lens--lens coupling between the main deflector and the perturbers. A point regarding notation: we later denote the real component of any shear with $\gamma_1$, and the imaginary component with $\gamma_2$, with the magnitude $|\gamma| = \sqrt{\gamma_1^2 + \gamma_2^2}$.

Measuring the three individual shears $\gamma_{\rm os}$, $\gamma_{\rm ds}$ and $\gamma_{\rm od}$ from a strong lensing image is impossible, since they are degenerate with each other and with parameters of the model used to describe the gravitational potential of the main deflector, $\psi(\bm{\theta})$ \citep{Hogg:2022ycw}; if the iso-potential contours of the lens mass are elliptical, there is an exact degeneracy between the ellipticity and the foreground shear $\gamma_{\rm od}$. This is due to the fact that we have no access to the true, unlensed source position $\bm{\beta}$ or its unlensed morphology, meaning that up to linear transformations of the source, image positions are preserved; this is called the source-position transformation degeneracy \citep{Schneider:2013wga}, which is a generalisation of the mass-sheet degeneracy \citep{Falco1985}. 

However, this degeneracy can also be exploited in order to define a shear quantity which is measurable and free from degeneracies with lens model properties. This can be done by arbitrarily fixing a definition for $\bm{\beta}$, analogous to lifting a gauge freedom by fixing the gauge. It was shown by \cite{Fleury:2021tke} that a minimal model for LOS perturbations can be found by multiplying \autoref{eq:lens_eqn_los} by $\amplification\e{od}\amplification^{-1}\e{ds}$, yielding
\begin{equation}
\tilde{\bm{\beta}} = \amplification\e{LOS} \bm{\theta} - \frac{\dd \psi_{\rm eff}}{\dd \bm{\theta}} \ , \label{eq:lens_eqn_minimal}
\end{equation}
with the transformed source position $\tilde{\bm{\beta}} \equiv \amplification\e{od}\amplification\e{ds}^{-1}\bm{\beta}$, the LOS amplification matrix $	\amplification\e{LOS} \equiv \amplification\e{od} \amplification\e{ds}^{-1} \amplification\e{os}$ and the effective gravitational potential $\psi\e{eff}(\bm{\theta}) \equiv \psi(\amplification\e{od}\bm{\theta})$. \autoref{eq:lens_eqn_minimal} describes a main lens with potential $\psi\e{eff}$ and external tidal perturbations, $\amplification\e{LOS}$, located in the same plane, and is equivalent to \autoref{eq:single_plane_lens} up to a source-position transformation. This means that we can consider $\psi\e{eff}$ as representative of the real physical potential of the main deflector, $\psi$, up to the same source-position transformation. The minimal model for LOS perturbations thus contains five meaningful parameters: the real and imaginary components of the foreground and LOS shears, and the LOS rotation term. The convergence parameters play no role, thanks to the mass-sheet degeneracy. 

It was argued by \cite{Fleury:2021tke} that the LOS shear, the combination $\gamma_{\rm LOS} \equiv \gamma_{\rm od} + \gamma_{\rm os} - \gamma_{\rm ds}$ which arises from the  $\amplification\e{LOS}$ amplification matrix in the minimal model, is not degenerate with the ellipticity of the main lens at leading order. This was further investigated by \cite{Hogg:2022ycw}, who fit different models to a \textit{Hubble} Space Telescope-like catalogue of mock strong lensing images, showing that $\gamma_{\rm LOS}$ is recoverable and non-degenerate with other lens model parameters, provided the lens mass model has a complexity equivalent to that with which the mock images were generated. The current work takes the logical next step, providing the first measurement of the LOS shear in real observational data and beginning to explore potential systematics in that measurement due to the shear and lens mass models, laying the groundwork for the cosmological constraints described above.

\section{Data and methodology}\label{sec:methods}
In this section we describe the image data used, the components of our lens modelling and the method used to fit lens models to the images.

\subsection{The strong lens sample}
The Sloan Lens ACS (SLACS) strong lens catalogue is a large, well-studied sample of lenses, making it an ideal testbed for a first measurement of the LOS shear. Candidate lenses were identified spectroscopically in Sloan Digital Sky Survey (SDSS) data  and followed up with \textit{Hubble} Space Telescope (HST) observations \citep{Bolton2006}. The lenses we model in this work were imaged either with the Advanced Camera for Surveys (ACS) in the F555W filter or with the Wide Field and Planetary Camera 2 (WFPC2) in the F606W filter. The imaging data were obtained under the HST General Observers programmes 10494
(PI: Koopmans), 10798 (PI: Bolton), 10886 (PI: Bolton), and 11202 (PI: Koopmans).

A subset of 50 representative strong lenses from the SLACS catalogue were selected for study by \cite{Shajib2021}, with the criteria that the chosen lenses should not have satellite galaxies or complex source morphologies, in order to allow an automated modelling procedure to be applied. Furthermore, lenses were selected that have imaging data in the F555W and F606W filters, as the deflector light tends to be fainter in these bands compared to the lensed source light, which makes it easier to deblend the source and deflector light during lens modelling. Lastly, no disc-like lenses were selected. In their study, 23 of the 50 were modelled successfully.

The WFPC2 images were reduced for the original SLACS analysis; ACS images were reduced by \cite{Shajib2021} using the \texttt{Astrodrizzle} package \citep{Avila2015}, and the point-spread function (PSF) for each filter and camera combination was computed using \texttt{TinyTim} \citep{Krist2011}. In this work, we make use of these imaging data and PSFs, modelling the same 23 lenses as \cite{Shajib2021}. These lenses are listed in \autoref{tab:lenses}. We further extend our sample to 45 lenses in \cite{Hogg:2025asw}.

\subsection{Mass and light models}\label{subsec:models}
We model the mass of the main deflector using the elliptical power law (EPL) profile of \cite{Tessore:2015baa}. The convergence of this profile is given by
\begin{equation}
    \kappa(x, y) = \frac{3 - \gamma^{\rm EPL}}{2} \left(\frac{\theta_{\rm E}}{\sqrt{q x^2 + y^2/q}}\right)^{\gamma^{\rm EPL}-1},
\end{equation}
where $\theta_{\rm E}$ is the Einstein radius, $\gamma^{\rm EPL}$ is the slope of the power law describing the three-dimensional mass distribution and $q$ is the axis ratio of the ellipse. The EPL profile is a relatively simplistic description of the lens mass, and suffers from known systematics in the absence of additional complexity \citep{Schneider2013, Cao2022}. Nonetheless, it has been widely used in relevant literature, e.g. \cite{Etherington:2022nzt, Etherington:2023yyh}, and thus serves as a foundational model for comparisons with such works, and one to which additional flexibility can be added.  

The Einstein radius of each lens is not included in our Markov chain Monte Carlo (MCMC) parameter inference -- which we describe in more detail in \autoref{subsec:dolphin} -- as a sampled parameter. Instead, we fix $\theta_{\rm E}$ of each lens to its best fit value as reported by \cite{Shajib2021} (see Table 1 of that work). The motivation for this choice is to improve the computational efficiency of the MCMC. We note that this choice may lead to some loss of uncertainty in the remaining parameters, but since $\theta_{\rm E}$ typically has the smallest inherent uncertainty of all parameters in the model, the change is not significant.

We model the light of the main deflector with two elliptical S\'ersic profiles \citep{Sersic1963,Sersic1968}. The surface brightness of a S\'ersic profile is given by
\begin{equation}
    I(x, y)  = I_{\rm eff} \; {\rm exp} \left[ k -k \left( \frac{\sqrt{q x^2 + y^2/q}}{R_{\rm eff}} \right)^{1/n_{\rm S}} \right],
\end{equation}
where $R_{\rm eff}$ is the effective radius of the profile, $I_{\rm eff}$ is the surface brightness at that radius, $q$ is the axis ratio, $n_{\rm S}$ is called the S\'ersic index and $k$ is a normalising constant such that $R_{\rm eff}$ is the half-light radius.

We use two S\'ersic profiles rather than one to model the main deflector's light, as it has been shown, when fitting the specific lens RXJ1131-1231, that a single S\'ersic profile results in significant residual differences between observation and model in the centre of the image \citep{Claeskens:2006ky, Suyu2013}. We constrain the profiles so that their centres share the same position in the lens plane, and furthermore, fix the S\'ersic indices of the two profiles to one and four, $n_{\rm S} = 1$ and $n_{\rm S} = 4$, representing an exponential and a de Vaucouleurs profile respectively \citep{deVaucouleurs1948}. 

We model the light of the source with a further two light profiles: an elliptical S\'ersic profile plus a set of shapelets. Shapelets are two-dimensional Gauss--Hermite polynomials, and thanks to their forming a complete basis set, any image can be decomposed into a linear combination of these polynomials \citep{Refregier:2001fd}. They have been used to measure weak lensing shear in galaxy images by \cite{Refregier:2001fe}, and for numerous source reconstructions from strong lensing images, beginning with \cite{Birrer:2015rpa}.

We here summarise the shapelets formalism as presented in \cite{Refregier:2001fd}. The dimensionless basis functions, or shapelets, are defined in one dimension as
\begin{equation}
	\phi_n(x) \equiv \left[2^n \pi^{\frac12}n! \right]^{-\frac12} H_n(x) \; e^{-\frac{x^2}{2}},
\end{equation}
where $H_n(x)$ is the Hermite polynomial of order $n$. A dimensionful characteristic scale $\beta$ can be introduced,
\begin{equation}
B_n(\mathrm{x}; \beta) \equiv \beta^{-\frac12} \phi_n(\beta^{-1} \mathrm{x}),
\end{equation}
where $\mathrm{x}$ has units of length, to preserve the dimensionlessness of the argument of $\phi_n (x)$. Then, the intensity of a light source $f(\mathrm{x})$ can be decomposed into the basis functions
\begin{equation}
	f(\mathrm{x}) = \sum_{n=0}^{\infty} f_n B_n(\mathrm{x}; \beta),
\end{equation}
where $f_n$ are the shapelet coefficients. To describe the intensity of light across an image, this formalism can be extended to two dimensions, $\mathbf{x} = (\mathrm{x}_1, \mathrm{x}_2)$ and $\mathbf{n} = (n_1, n_2)$, in which case $\phi_{\mathbf{n}}(\mathbf{x}) \equiv \phi_{n_1} (x_1) \phi_{n_2}(x_2)$ and $B_{\mathbf{n}} (\mathbf{x}; \beta) \equiv \beta^{-1} \phi_{\mathbf{n}} (\beta^{-1} \mathbf{x})$. 

In practice, the parameters of our set of shapelets are the maximum order for the  Hermite polynomial $n_{\rm max} \geq n_1 + n_2$, which in turn determines the number of shapelets in the set, $N_{\rm shapelets} = (n_{\rm max} + 1)(n_{\rm max}+2)/2$, and the characteristic scale $\beta$. In our analysis, we fix $n_{\rm max}$ to a specific value for each lens, starting from the values used in \cite{Shajib2021} and updating some in a trial-and-error process. The values used are listed in \autoref{tab:lenses}. The scale parameter $\beta$ is inferred by the MCMC procedure described below. The centre of the set of shapelets is constrained to be at the same position in the source plane as the centre of the  elliptical S\'ersic profile that makes up the rest of the source light model.

Gaussian shapelets suffer from the limitation that the \nth{0} order function -- the Gaussian -- is not a realistic representation of the light profile of a typical galaxy \citep{Ngan:2008rg}. This can lead to over-fitting, particularly in the outskirts of the galaxy. As an alternative, \cite{Berge:2019nyt} proposed one and two-dimensional exponential shapelets, based on wave functions of the hydrogen atom, which they argued are more suitable for describing galaxy shapes in astronomical imaging, amongst other applications. In this work, we tested the exponential shapelets for source reconstruction on five of our sample of lenses, finding generally poorer fits. For this reason, and for the reason of preserving a fair comparison between our results and those of \cite{Shajib2021}, we continued to use the Gaussian shapelets for our source reconstructions. A more rigorous exploration of the use of exponential, or other forms of shapelets, is left to a future work.

\subsection{Shear models}
We use two different models for the shear of the strong lensing images. The first is the minimal LOS shear model, described by \autoref{eq:lens_eqn_minimal}. In this case, the shear model has five parameters: the LOS shear parameters $\gamma_1^{\rm LOS}$ and $\gamma_2^{\rm LOS}$, the foreground shear parameters $\gamma_1^{\rm od}$ and $\gamma_2^{\rm od}$ and the LOS rotation parameter $\omega_{\rm LOS}$. 

In \cite{Hogg:2022ycw}, it was shown that when fitting lens models to mock data, the original image could be recovered even when the foreground shear was neglected in $\psi_{\rm eff}$. In the second shear model we consider, we omit the explicit appearance of $\gamma_{\rm od}$ in the effective main-lens potential. The foreground shear term physically corresponds to the effect of post-Born corrections in the main deflection -- in the presence of foreground perturbations, light rays do not trace to the same position in the observer's plane as they would if those perturbations were absent. 

Thus, the omission of the foreground shear must be understood as the assumption that such post-Born corrections are degenerate enough with the parameters of the main lens to be absorbed in them; we do not assume that foreground shear, nor its post-Born effect on the main deflector, are negligible. Practically speaking, this means that we fix $\gamma_1^{\rm od} = \gamma_2^{\rm od} = 0$ in our analysis using this model, which we refer to as the `no foreground shear model' throughout the rest of this paper. 

These models are both mathematically distinct from the notion of `external' or `residual' shear which is widely used in the literature (\cite{Tan2024}; see \cite{Shajib2024} for a discussion of this terminology). The LOS formalism can be reduced to the external shear model by fixing both the foreground shear and LOS rotation to zero, $\gamma_{\rm od} = \omega_{\rm LOS} = 0$, which leaves the shear to be described by just a single complex number, with the real and imaginay components $\gamma_1$ and $\gamma_2$.

We note that in practice, what is actually measured in each of these cases is the `reduced shear', $\mathrm{g}_{\rm ab} = \gamma_{\rm ab}/(1 - \kappa_{\rm ab})$. The convergence $\kappa_{\rm ab}$ is typically small (and in fact, unmeasurable in strong lensing due to the mass-sheet degeneracy), $\kappa_{\rm ab} \ll 1$, meaning that we can approximate $\mathrm{g}_{\rm ab} \approx \gamma_{\rm ab}$. Throughout the rest of this work, we will therefore refer to what is technically `reduced shear' as `shear'.

\subsection{The automated modelling procedure}\label{subsec:dolphin}
Before beginning the modelling, we apply a custom annular mask to each of our 23 selected SLACS lenses, with the inner and outer radii selected manually for each lens (typically around $0.4''$ and $3.0''$ respectively) masking out the central and outer regions of the images. We also mask the light from any prominent foreground object in the field of view. This masking is an alternative to the lens light removal, for example via a multi-Gaussian expansion, that is performed in other analyses, e.g. \cite{He:2024udi}. However, when the mask is in place, we still model the remaining lens light, allowing us to infer the uncertainties in the model parameters. This could not be done if the lens light is removed before any modelling step.

We then fit lens models to the images using the \lenstronomy package \citep{Birrer:2018xgm, Birrer2021}, via the \dolphin software \citep{Shajib2021,Tan2024, Shajib:2025bho}, which provides an automated modelling pipeline to uniformly model a large sample of lenses. The modelling has two stages: first, a particle swarm optimisation (PSO) \citep{Eberhart1995} is used to locate the maximum likelihood value for each parameter in the model, i.e. the lens mass, lens light and source light parameters.

The \dolphin package provides a recipe to increase the computational efficiency of the optimisation step, in which the lens mass, lens light and source light model parameters are individually optimised whilst the rest of the parameters are kept fixed, before all the model parameters are optimised together. Full details of this optimisation recipe can be found in \cite{Shajib2021}. 

In the second stage of the parameter inference, an MCMC method is used to sample from the posterior distribution formed by the likelihood and the user-defined priors for each parameter. In this work we use the affine-invariant ensemble sampling algorithm provided by the \texttt{emcee} package \citep{GoodmanWeare, ForemanMackey2013}; the ensemble of walkers is initialised in a small Gaussian ball around the maximum likelihood value found by the PSO. Our chosen priors are listed in \autoref{tab:priors}. 

The positions of the walkers in the ensemble are updated using a random choice of either the differential evolution (DE) \citep{Nelson2014} or DE `snooker' moves \citep{terBraak2008}, with a respective probability of 80\% and 20\%, which is more efficient\footnote{See the \texttt{emcee} documentation: \url{https://emcee.readthedocs.io/en/stable/tutorials/moves}.} for sampling from mildly multimodal distributions than the standard `stretch' move of \cite{GoodmanWeare}.

Multimodality may be encountered in our inference when foreground shear and the ellipticity of the EPL profile conspire to produce a given image with interchangeable orientation angles; for example, the combination of $|\gamma_{\rm od}|$ at an angle of $0\degree$ East of North and an EPL ellipticity of $|e_{\rm EPL}| \equiv |\gamma_{\rm od}|$ at $45\degree$ East of North is indistinguishable in effect from the same quantities with their angles exchanged. This can result in bimodal marginalised posterior distributions for the components of these complex variables. The effect is simply a manifestation of the constructed degeneracy between $\gamma_{\rm od}$ and lens mass ellipticity in the minimal LOS model.

Besides the priors on the individual parameters, \dolphin offers the user the possibility to add other prior constraints on the parameters. In particular, a limit can be imposed on the maximum difference between the orientation of the lens mass and lens light ellipticities, as well as a limit on the ratio of the axis ratios of the lens mass and lens light. These priors reflect the physical expectation that the underlying matter distribution in a galaxy will trace the visible light. 

However, this is only true for objects with a relatively insignificant foreground; if a non-negligible foreground is present, the apparent orientation of the lens light will generally differ from both the apparent and intrinsic orientation of the mass model ellipticity. In this case, a prior rigidly joining lens light and lens mass ellipticity is no longer physically well-motivated, and the degeneracies present between foreground shear, $\gamma_{\rm od}$, and the lens mass ellipticity will lead to biased or simply incorrect parameter inference across the model; see \cite{Johnson:2024hvl} for further discussion of foreground biases in strong lensing. Given the entanglement between foreground effects and the lens model, this type of prior should not be used when attempting to infer LOS shear. 

\section{Results and discussion}\label{sec:results}
The above-described automated modelling procedure allowed us to successfully model all 23 lenses in the \cite{Shajib2021} sample. Our results are presented after post-processing the chains, including the removal of burn-in and smoothing of the marginalised posterior distribution histograms.

\subsection{Shear is measurable in the minimal LOS model}\label{subsec:min}

\begin{figure*}
	\centering
	\includegraphics[width=\textwidth]{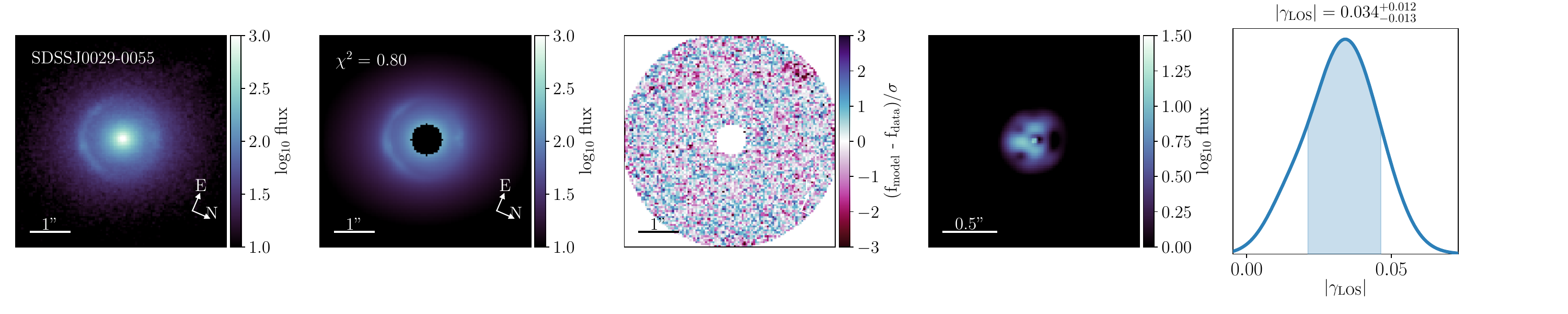}\\
	\includegraphics[width=\textwidth]{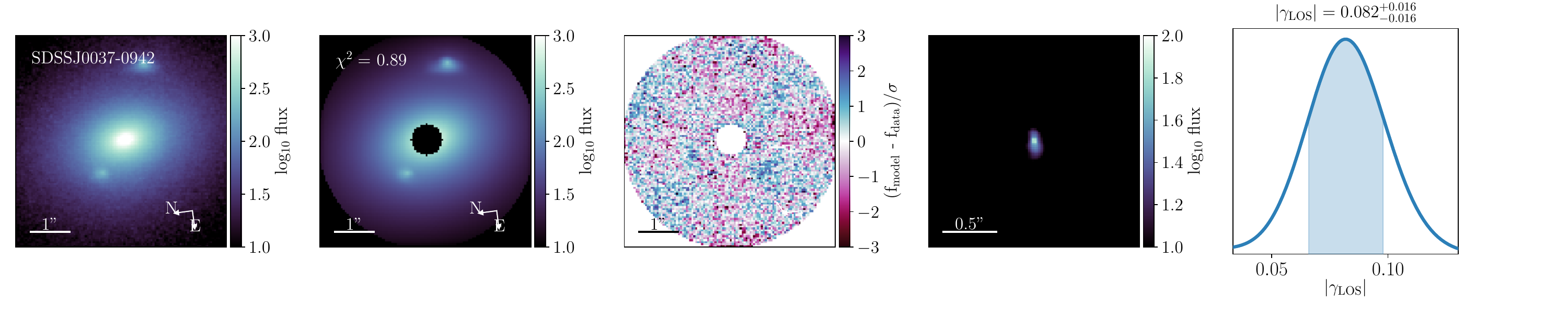}\\
    \includegraphics[width=\textwidth]{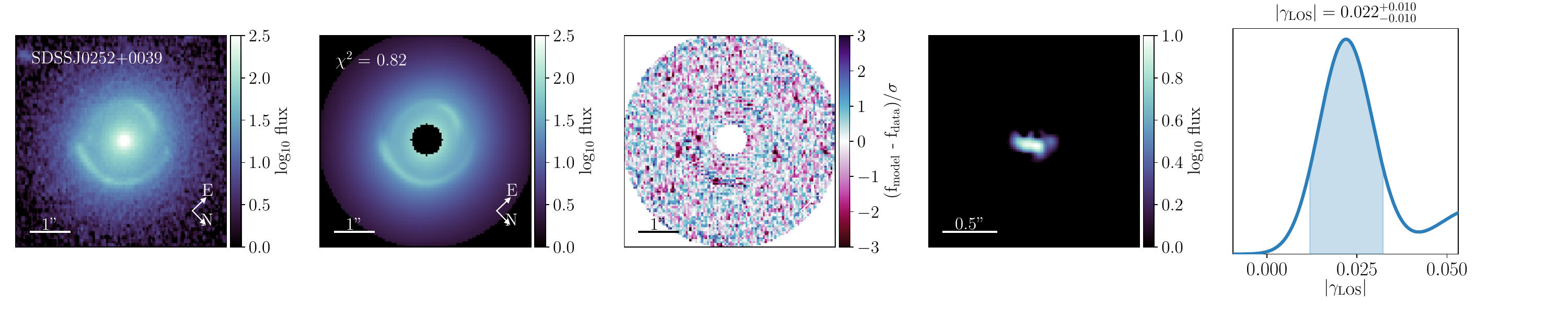}\\
    \includegraphics[width=\textwidth]{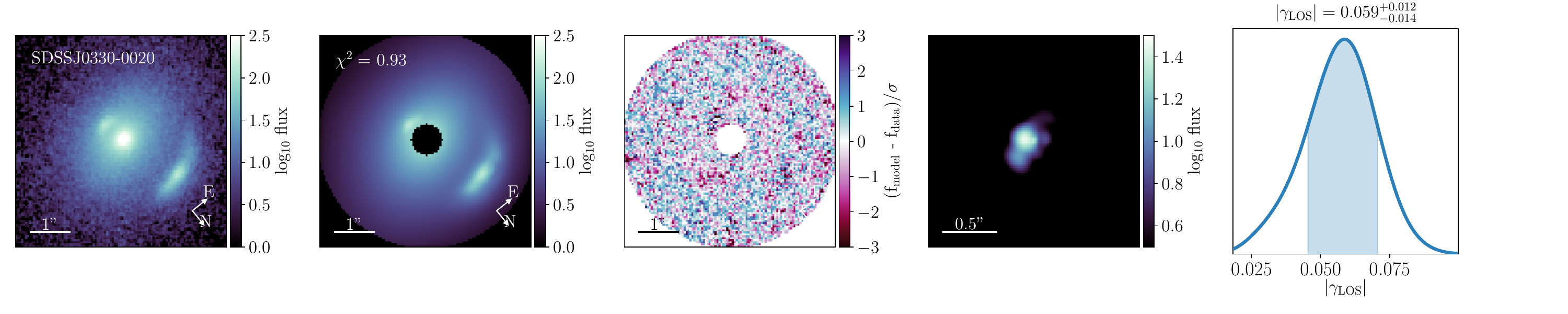}\\
    \includegraphics[width=\textwidth]{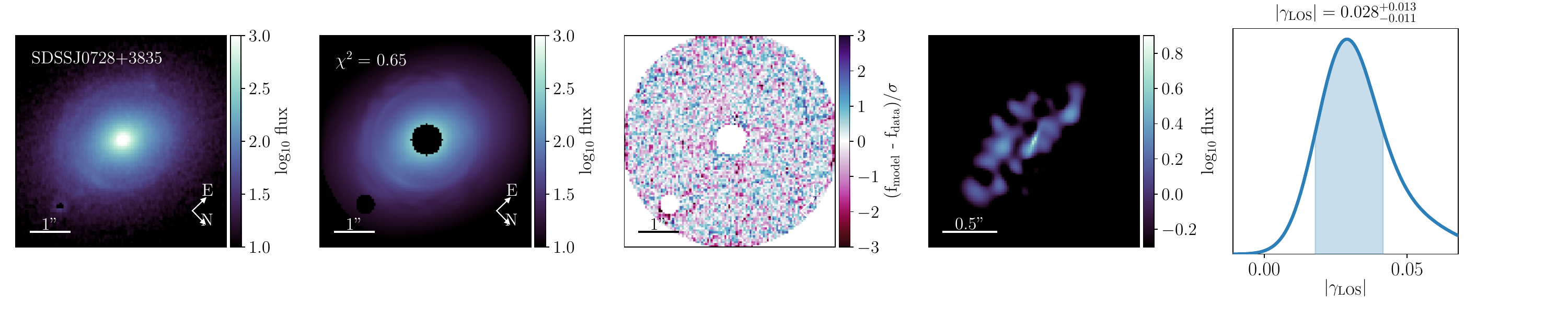}\\
    \includegraphics[width=\textwidth]{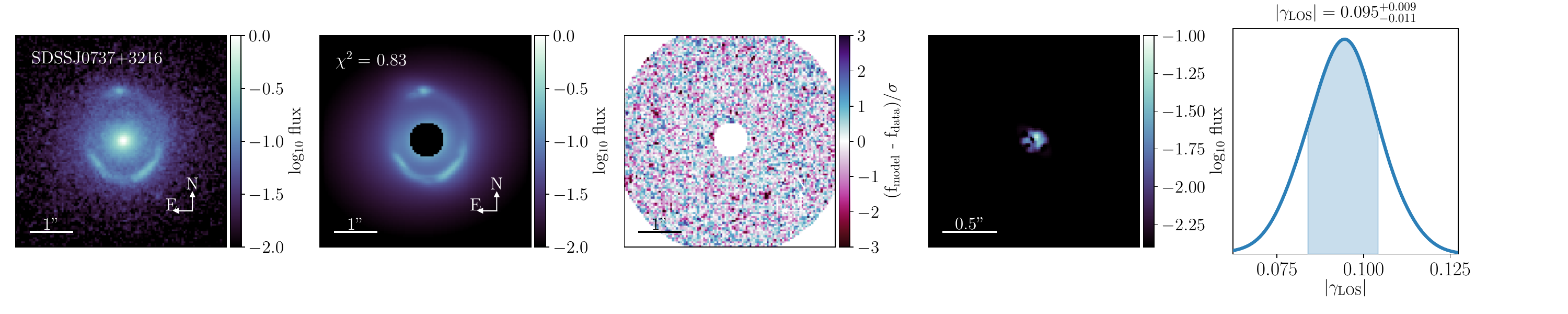}
	\caption{The first six lenses fit with the minimal model. From left to right, the panels show the single-band image data for each lens, our reconstruction of the image along with the reduced $\chi^2$ of the model, the residual difference between the image and the reconstruction, the reconstructed source and the one dimensional marginalised posterior distribution of the LOS shear magnitude, $|\gamma_{\rm LOS}|$. The shaded area is the $1 \sigma$ confidence interval.}
	\label{fig:min1}
\end{figure*}

\begin{figure*}
	\centering
	\includegraphics[width=\textwidth]{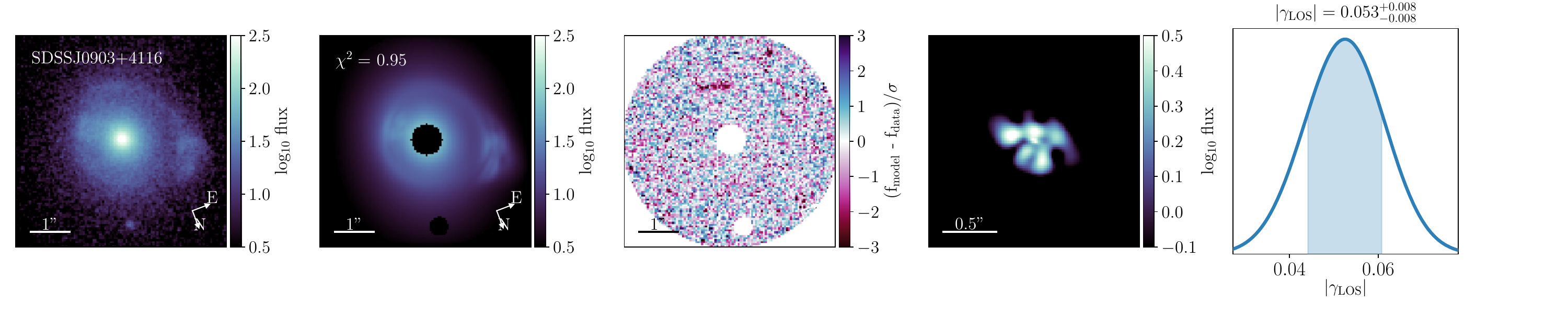}\\
	\includegraphics[width=\textwidth]{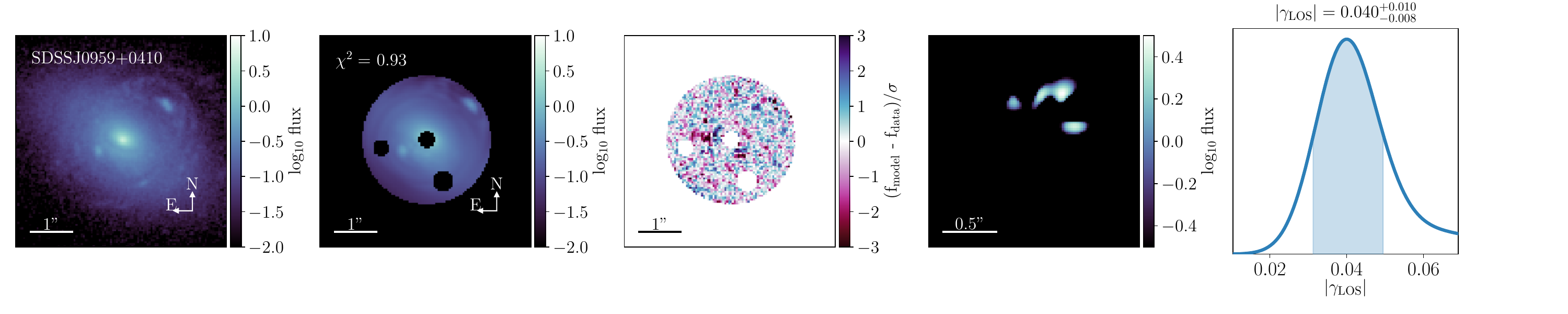}\\
    \includegraphics[width=\textwidth]{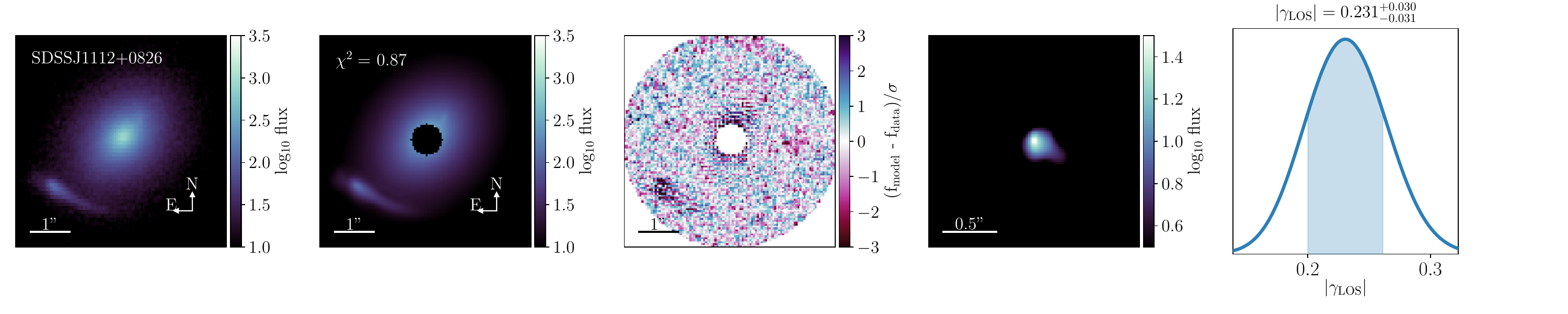}\\
    \includegraphics[width=\textwidth]{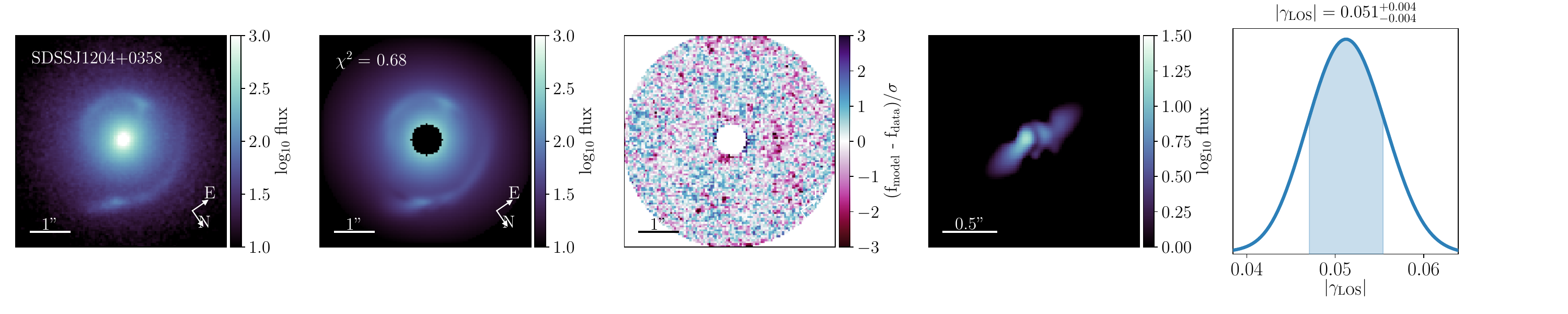}\\
    \includegraphics[width=\textwidth]{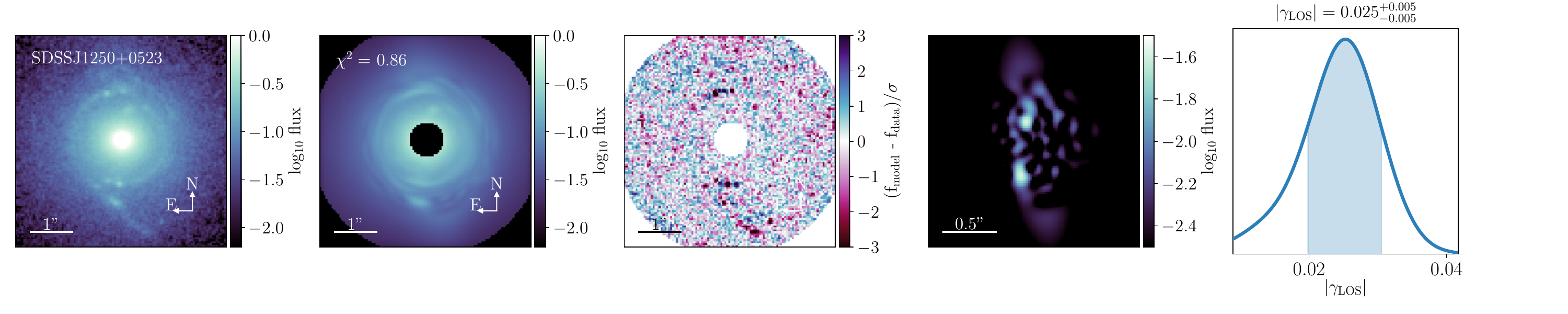}\\
    \includegraphics[width=\textwidth]{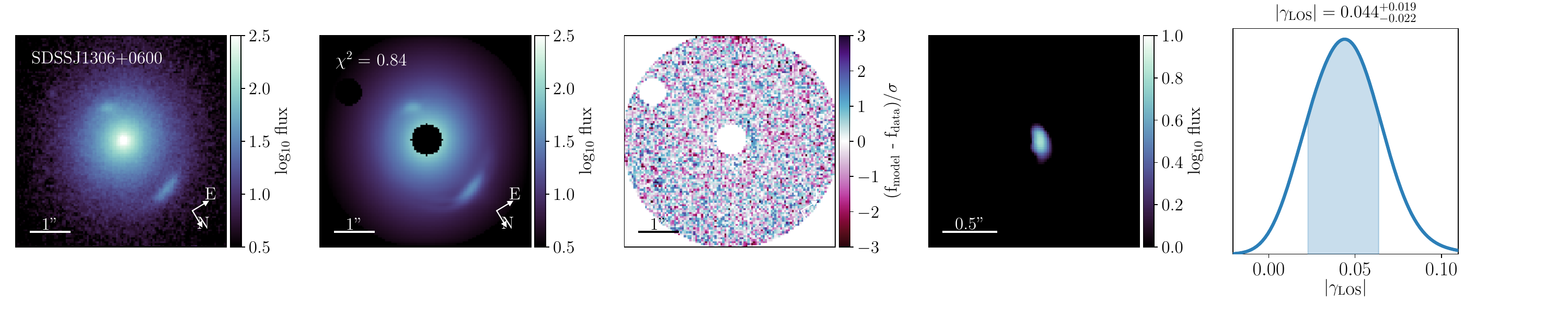}
	\caption{The next six lenses fit with the minimal model. The panels show the same information as in \autoref{fig:min1}.}
	\label{fig:min2}
\end{figure*}

\begin{figure*}
	\centering
	\includegraphics[width=\textwidth]{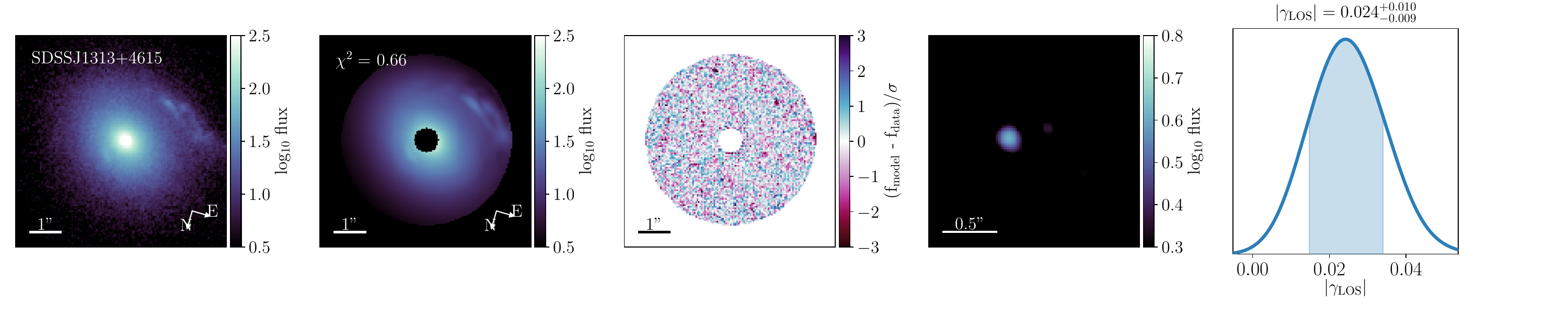}\\
	\includegraphics[width=\textwidth]{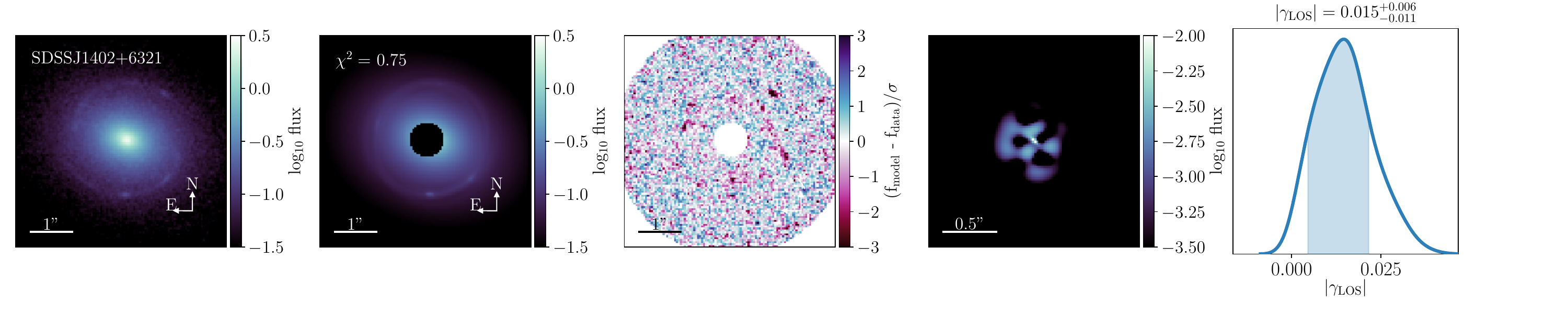}\\
    \includegraphics[width=\textwidth]{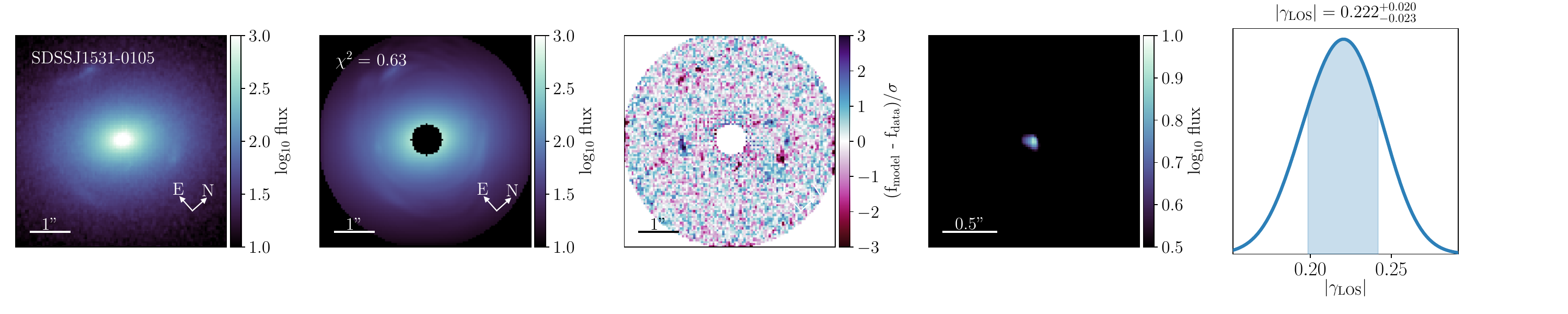}\\
    \includegraphics[width=\textwidth]{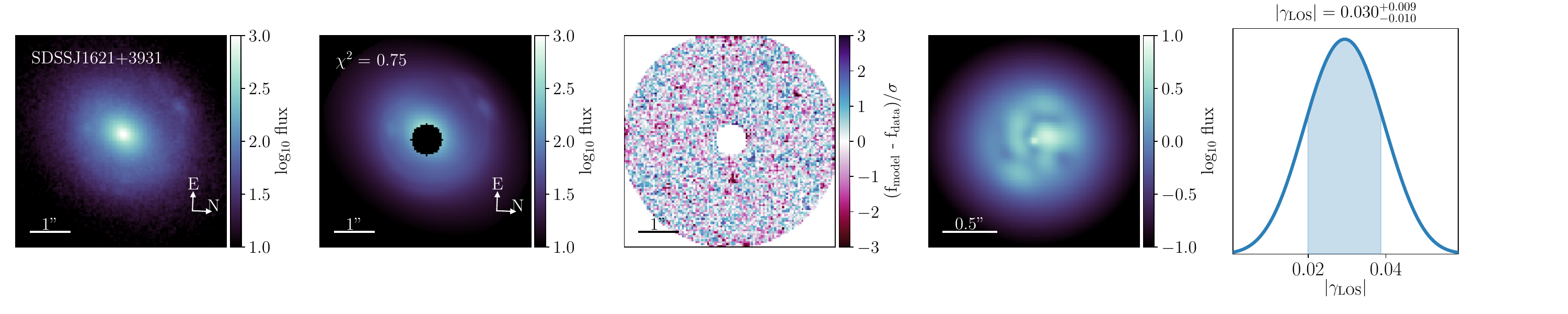}\\
    \includegraphics[width=\textwidth]{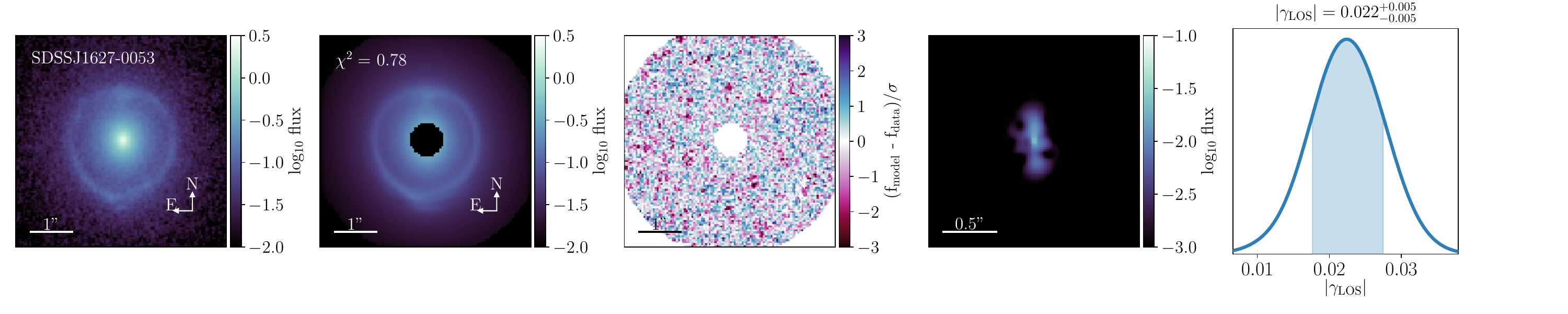}\\
    \includegraphics[width=\textwidth]{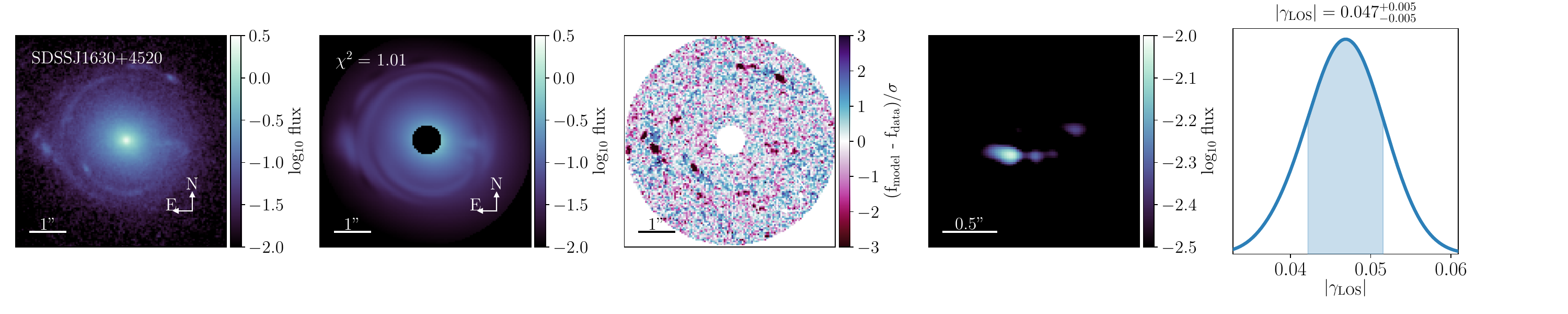}
	\caption{The next six lenses fit with the minimal model. The panels show the same information as in \autoref{fig:min1}.}
	\label{fig:min3}
\end{figure*}

\begin{figure*}
	\centering
	\includegraphics[width=\textwidth]{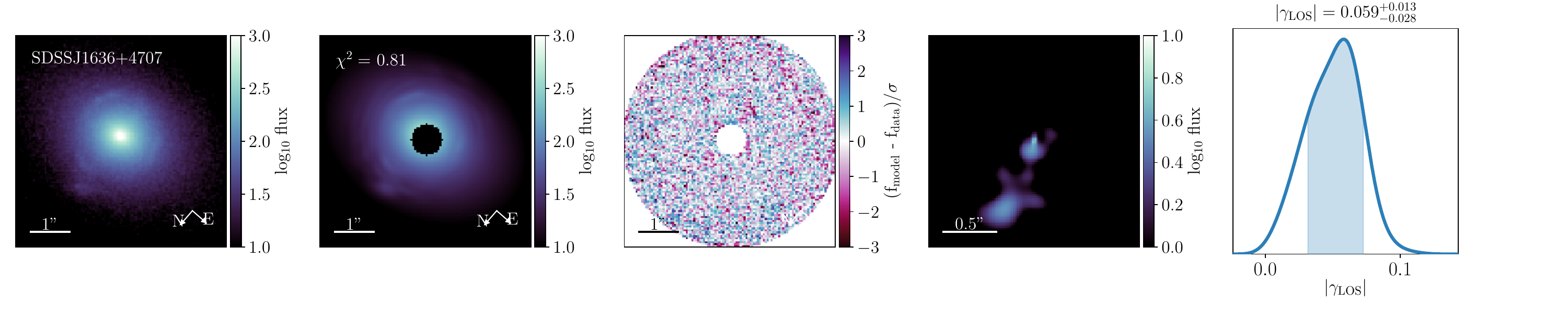}\\
	\includegraphics[width=\textwidth]{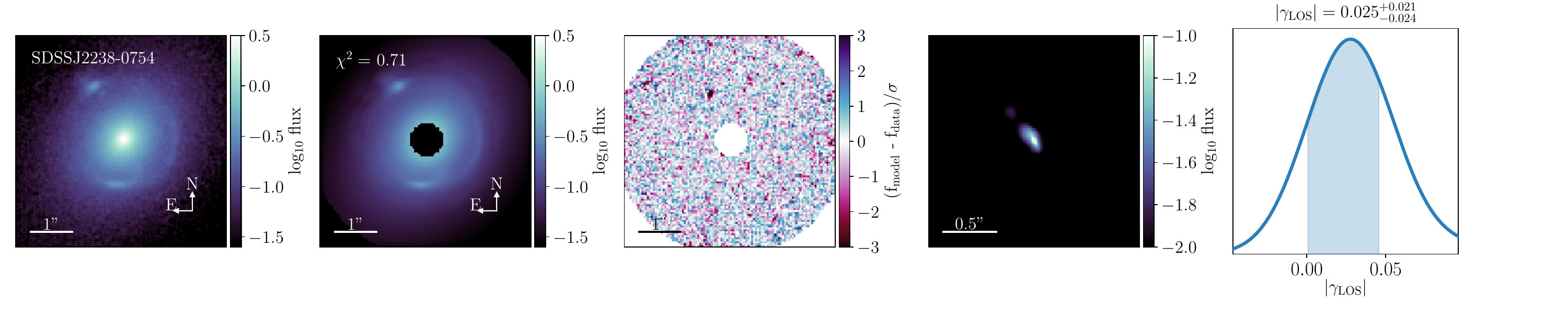}\\
    \includegraphics[width=\textwidth]{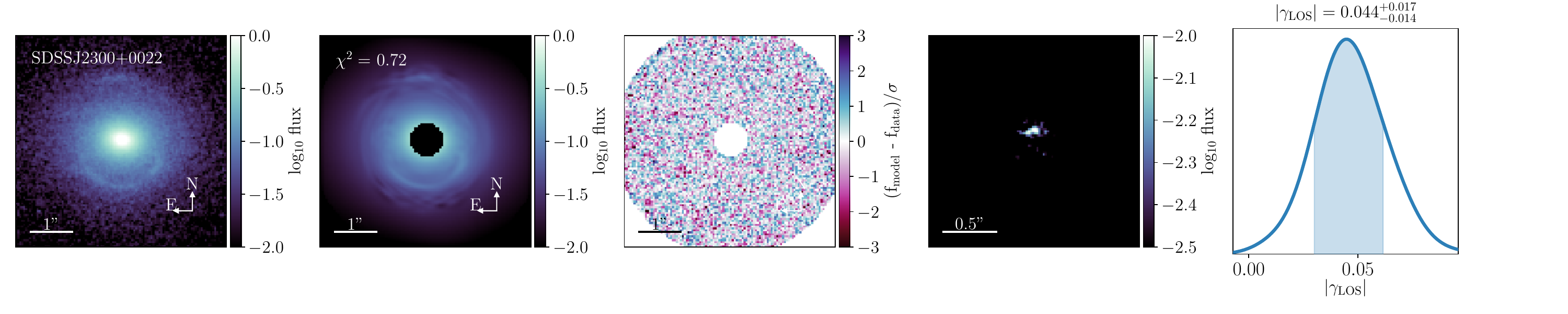}\\
    \includegraphics[width=\textwidth]{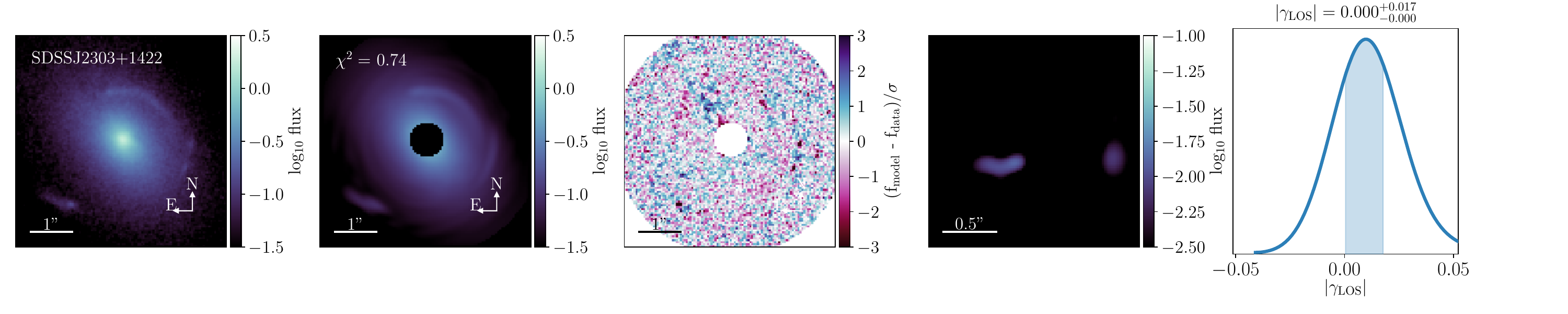}\\
    \includegraphics[width=\textwidth]{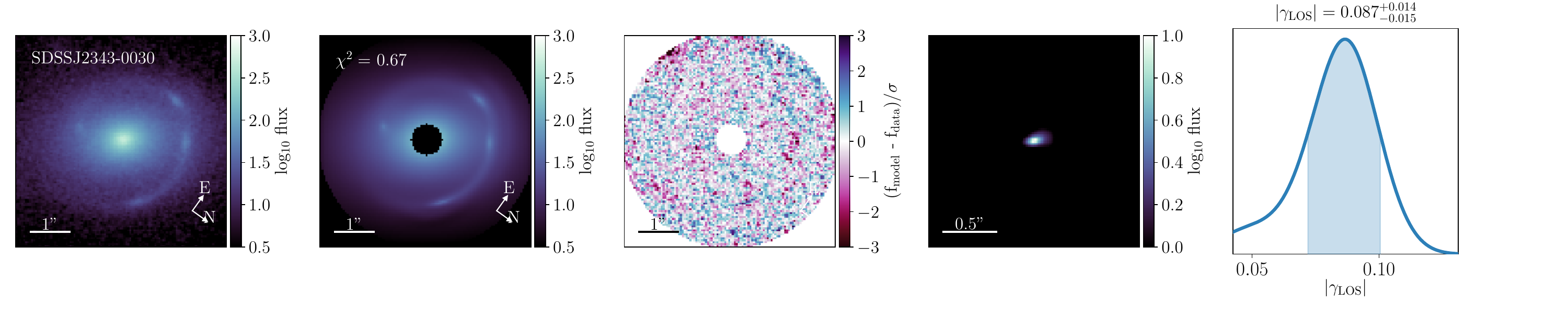}
	\caption{The final five lenses fit successfully with the minimal model. The panels show the same information as in \autoref{fig:min1}.}
	\label{fig:min4}
\end{figure*}

In \autoref{fig:min1}, \autoref{fig:min2}, \autoref{fig:min3}, and \autoref{fig:min4}, we show the results of our lens modelling with the minimal LOS shear model, and the other model components described above. The first column of these figures shows the image data for each strong lens; the second column shows the model reconstruction of the image and associated $\chi^2$ value; the third column shows the residual difference between the data and model; the fourth column shows the reconstruction of the source; and the fifth and final column shows the 1D marginalised posterior distribution of the LOS shear magnitude, $|\gamma_{\rm LOS}|$, for that system. The posterior samples for this quantity are computed directly from the posterior samples of the individual shear components, $\gamma_1$ and $\gamma_2$.

\begin{figure}
	\centering
	\includegraphics[width=0.49\textwidth]{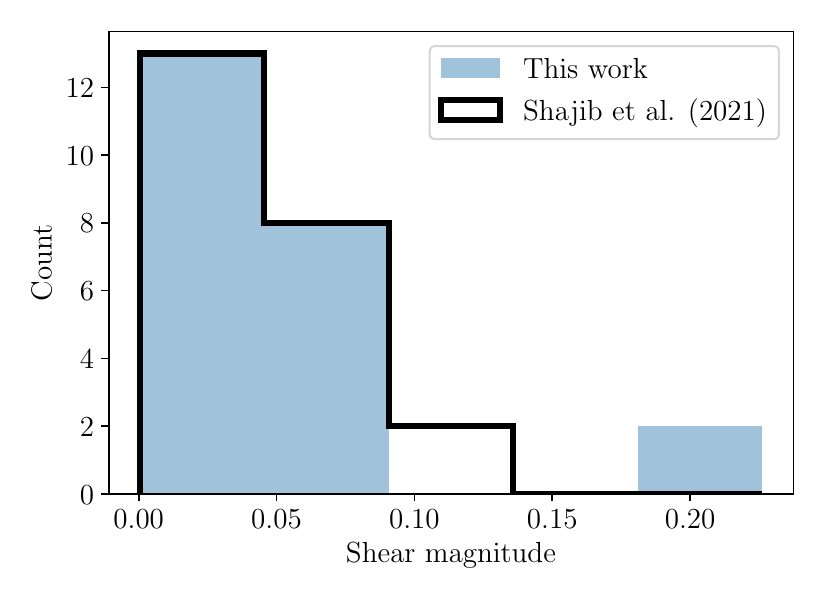}
	\caption{This figure shows the histograms of the $|\gamma_{\rm LOS}|$ values measured in this work (solid blue) and the external shear values measured in \cite{Shajib2021} (unfilled black).}
	\label{fig:shearhist}
\end{figure}

In \autoref{fig:shearhist}, we show the histogram of the magnitude of the LOS shears measured in our sample of 23 lenses in solid blue, compared with the histogram of the `external' shear found in the same lenses by \cite{Shajib2021}. From this figure, we can see that our shear measurements are very consistent with the previous study by \cite{Shajib2021}, with the exception of the two outliers SDSSJ1112+0826 at $|\gamma_{\rm LOS}| = 0.23 \pm 0.03$ and SDSSJ1531$-$0105 at $|\gamma_{\rm LOS}| = 0.21 \pm 0.03$.

We note that the majority of the reduced $\chi^2$ values for our fits are smaller than unity, indicative of over-fitting. During the trial-and-error experimentation described above to select a maximum order for the shapelets in the source model, $n_{\rm max}$, we noticed a trade-off between choosing relatively inflexible shapelets i.e. a small $n_{\rm max}$, leading to a reduced $\chi^2$ close to unity but prominent residuals between model and data; or choosing a more flexible set of shapelets, leading to a model that reproduces the data to the noise level, but has a $\chi^2$ on average below unity. We chose to prioritise the reduction of residuals over larger $\chi^2$ values. In future work, a fully Bayesian model comparison approach will be taken.

Furthermore, some of our reconstructed sources show `ringing' features characteristic of Gaussian shapelets. Exponential shapelets may mitigate this issue, but, as described in \autoref{subsec:models}, we found generally poorer fits i.e. strong residuals between model and data, when fitting a subset of these lenses with exponential shapelets.

We note that bimodality is present in the joint posteriors of the lens mass ellipticity and the foreground shear for 6 of the 23 lenses. However, this is not a particular issue for our inference, for the following reasons. Firstly, we assert that all chains are converged, by examining the trace plots of the walkers and computing the autocorrelation times for the chains. Secondly, the bimodality is not present in the posteriors for $\gamma_{\rm LOS}$, our main parameter of interest, thanks to the minimal model. Thirdly, even if bimodality were present in parameters of interest, the kernel density estimation method by which we infer posterior means and uncertainties (via the Python package \texttt{chainconsumer} \citep{Hinton2016}) will overestimate uncertainty for weakly bimodal posteriors, and fail for posteriors with two or more highly distinct modes. We are therefore never at risk of presenting a falsely \textit{under}estimated uncertainty on the LOS shear due to bimodality, meaning that this is the maximally conservative approach.

\subsection{The measured shear is larger than expected}\label{subsec:largeshear}

Across the set of 23 lenses, we measure the LOS shear with a mean magnitude of $0.056 \pm 0.013$. \cite{Etherington:2023yyh} showed that the large values of \textit{external} shear measured from strong lensing are inconsistent with expectations from weak lensing measurements, both in mock data and in observations of $\gamma\e{os}$ in the same regions of the sky as the strong lenses, a study which was conducted using a sample of the SLACS lenses which overlaps with ours (13 lenses in common). We are able to make an independent confirmation of this finding for LOS shear using $N$-body simulations.
	
We use the \texttt{RayGalGroup} suite of relativistic $N$-body simulations, whose publicly available data contains weak lensing maps that include post-Born effects, magnification bias and redshift-space distortions \citep{2019MNRAS.483.2671B,Rasera:2021mvk}. The maps provide convergence and shear measurements at a given set of redshifts, between which we can interpolate to compute $\gamma_{\rm os}$, $\gamma_{\rm od}$ and $\gamma_{\rm ds}$ shear values for any given deflector and source redshift, following the method presented in \cite{Johnson2025}. From these, we compute the expected $|\gamma_{\rm LOS}|$ for any line of sight in the simulation, and thus produce a distribution of expected $|\gamma_{\rm LOS}|$ shear magnitudes for either an ensemble of lines of sight, or for specific ($z_{\rm d}$, $z_{\rm s}$) pairs.

\begin{figure*}
	\centering
	\includegraphics[width=\textwidth]{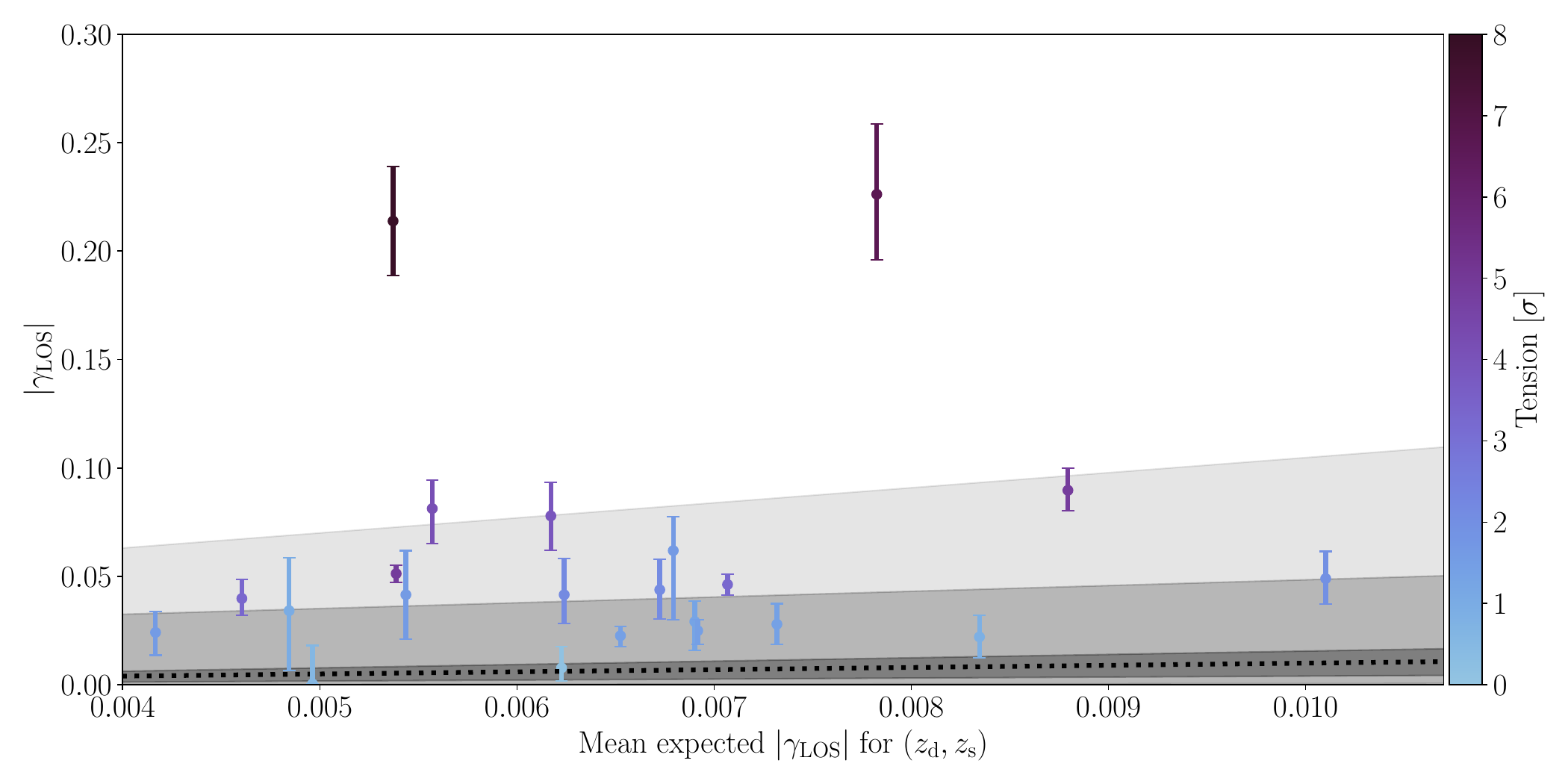}
	\caption{This figure shows the measured value of $|\gamma_{\rm LOS}|$ and the $1 \sigma$ error bars for each lens in our sample. The black dotted line shows the mean expected value of $|\gamma_{\rm LOS}|$  for that system's lens and source redshift, computed from the \texttt{RayGalGroup} simulations, as described in the main text. The grey shaded regions represent the $1$, $3$ and $5 \sigma$ uncertainties around this mean. The measured values are coloured by their tension in $\sigma$ from the mean, defined in \autoref{eq:tension}.}
	\label{fig:deltasigma}
\end{figure*}

In \autoref{fig:deltasigma}, we show our measured LOS shear value in each system against the mean expected $|\gamma_{\rm LOS}|$ for the deflector and source redshift of that observed system, obtained from the simulations. The black dotted line shows the mean, whilst the grey bands represent the $1$, $3$ and $5\sigma$ uncertainties. The measured points are coloured according to their tension with the simulation. The tension between two quantities $a$ and $b$ is defined as follows,

\begin{equation}
    T_{ab} = \frac{|x_a - x_b|}{\sqrt{\rule{0pt}{2ex}\sigma_a^2 + \sigma_b^2}},
    \label{eq:tension}
\end{equation}
where $x$ are the best-fit values and $\sigma$ the standard deviations of the two quantities being compared.

From \autoref{fig:deltasigma}, we can see that the majority of measured  $|\gamma_{\rm LOS}|$ values are larger than the associated expected mean, with the largest tension between measurement and simulation sitting at $7.9 \sigma$ for lens SDSSJ1531$-$0105 (all numerical tension values are listed in \autoref{tab:lenses}). A further seven lenses have tensions of $3 \sigma$ or more with the expectation from the simulation. In the next subsection, we will explore one possible explanation for these unexpectedly large shear values.

\subsection{Shear does not compensate for unmodelled octupoles}\label{subsec:bodi}
Previous works have asserted that the large values of external shear measured in strong lensing images are the result of unmodelled complexity in the dominant lens, proposing that a fourth-order multipole, or octupole, in the lensing potential may be responsible \citep{Etherington:2023yyh}. This so-called `boxyness' or `diskyness' has also been observed in the isophotes of elliptical galaxies \citep{Bender1988,Stacey2024}  and produced in simulations \citep{Naab:1999ge}. In this subsection, we investigate the effect including octupoles in our lens model has on the inferred LOS shear.

A general multipolar lensing potential is given by
\begin{equation}
	\psi(\bm{\theta}) = \theta \frac{a_m}{1 - m^2} \cos[m(\varphi - \varphi_m)],
\end{equation}
where $\theta$ and $\varphi$ are the radial and angular coordinates of $\bm{\theta}$, $m$ is the order of the multipole, $a_m$ is its strength and $\varphi_m$ is the orientation angle of its main axis~\hbox{\citep{Keeton:2002qt,Xu:2014dda}}. In the case of the octupole, $m=4$, and the shape is called `boxy' when $\varphi_4 = \frac{\pi}{4}$ and `disky' when $\varphi_4 = 0$. To investigate the effect the inclusion of an octupole has on the LOS shear, we re-model our sample of lenses using the \texttt{EPL\_BOXYDISKY} profile in \lenstronomy, which adds a pure octupolar distortion on top of the elliptical power law profile \citep{VandeVyvere2021}. The source models for each lens are kept the same. We use the minimal LOS shear model.

\begin{figure*}
	\centering
	\includegraphics[width=\textwidth]{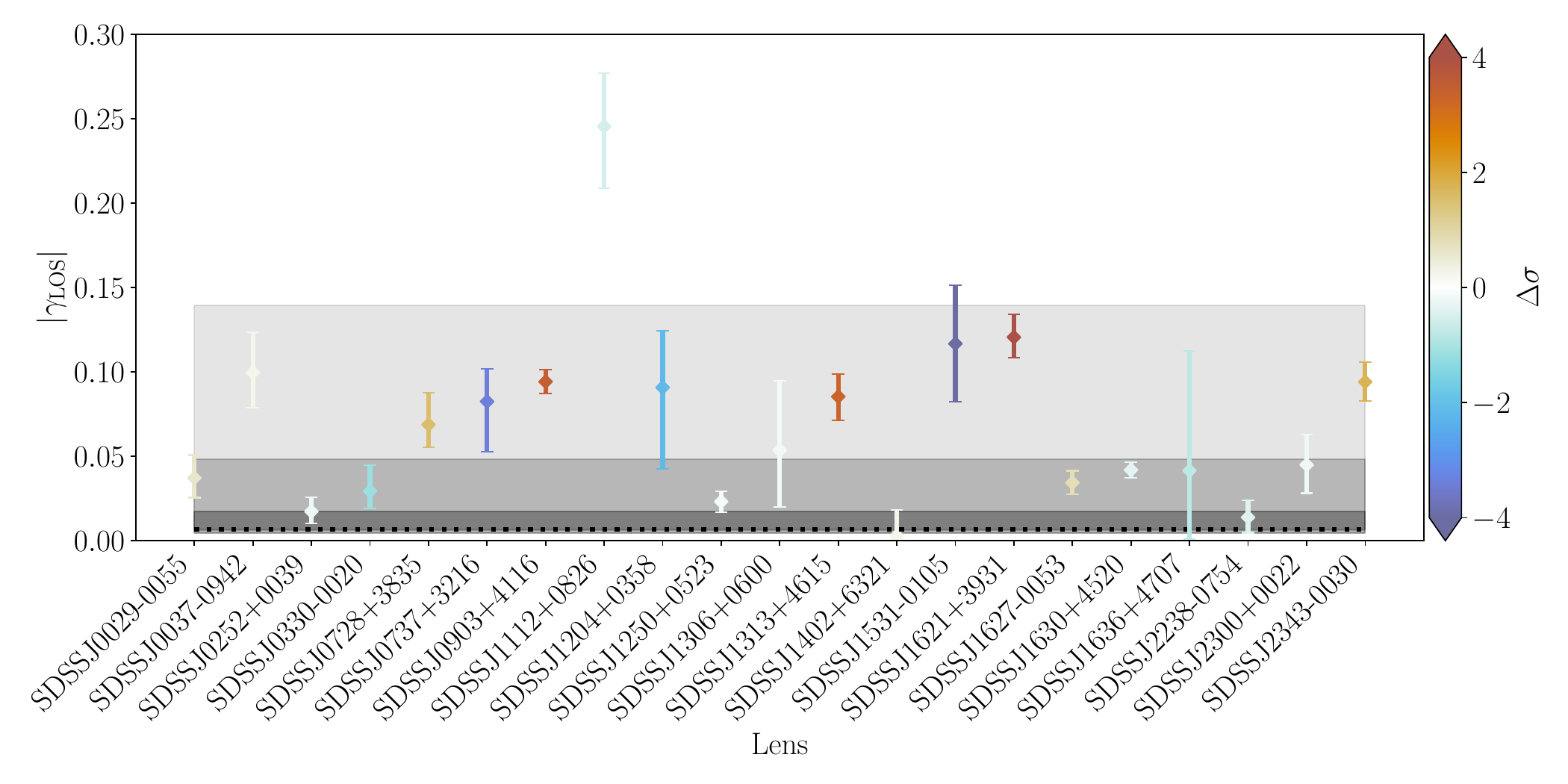}
	\caption{This figure shows the measured value of $|\gamma_{\rm LOS}|$ with the $m_4$ multipole included in the lens mass model. The black dotted line shows the overall mean expected value of  $|\gamma_{\rm LOS}|$, computed from the \texttt{RayGalGroup} simulations, as described in the main text, with the associated 1, 3 and $5 \sigma$ uncertainties around this mean given by the grey shaded bands. The measured values are coloured by the change in the tension between the minimal model and the minimal + octupole model.}
	\label{fig:specialbodi}
\end{figure*}

\autoref{fig:specialbodi} shows the measurement of $|\gamma_{\rm LOS}|$ in each lens with the addition of the octupole to the mass model. The systems are ordered by increasing deflector redshift on the $x$-axis. The black dotted line and grey bands represent the mean and $1$, $3$ and $5\sigma$ uncertainties on $|\gamma_{\rm LOS}|$ inferred from the \texttt{RayGalGroup} simulations, as described above. In contrast to \autoref{fig:deltasigma}, we plot the overall mean and uncertainties rather than those quantities computed for each lens, hence the absence of the redshift-dependent slope seen in that previous figure.

Furthermore, the measured points in this plot are coloured according to the change in the tension statistic compared to the result using the minimal LOS model without the additional octupole in the lens mass. All values are listed in \autoref{tab:lenses}. Points coloured in orange are in greater tension with the mean of the simulations compared to the measurement in the minimal model, whilst points coloured in blue are in lesser tension. We note that for SDSSJ0959+0410 and SDSSJ2303+1422, either one or both of the LOS shear components $\gamma_1$ and $\gamma_2$ were unconverged in the octupole model; we do not include these lenses in this plot.

In 14 of the remaining 21 lenses, the tension with the simulations changes by $1.5 \sigma$ or less following the addition of the octupole to the model; this is already sufficient to conclude that the inferred LOS shear does not \textit{systematically} decrease in magnitude when the octupole is included. Of the remaining seven lenses, four show an \textit{increase} in the tension in the range of $1.7$--$5.5 \sigma$, again supporting the conclusion that unmodelled octupoles are not artificially boosting the inferred shear in strong lenses.

\begin{figure*}
	\centering
	\includegraphics[width=\textwidth]{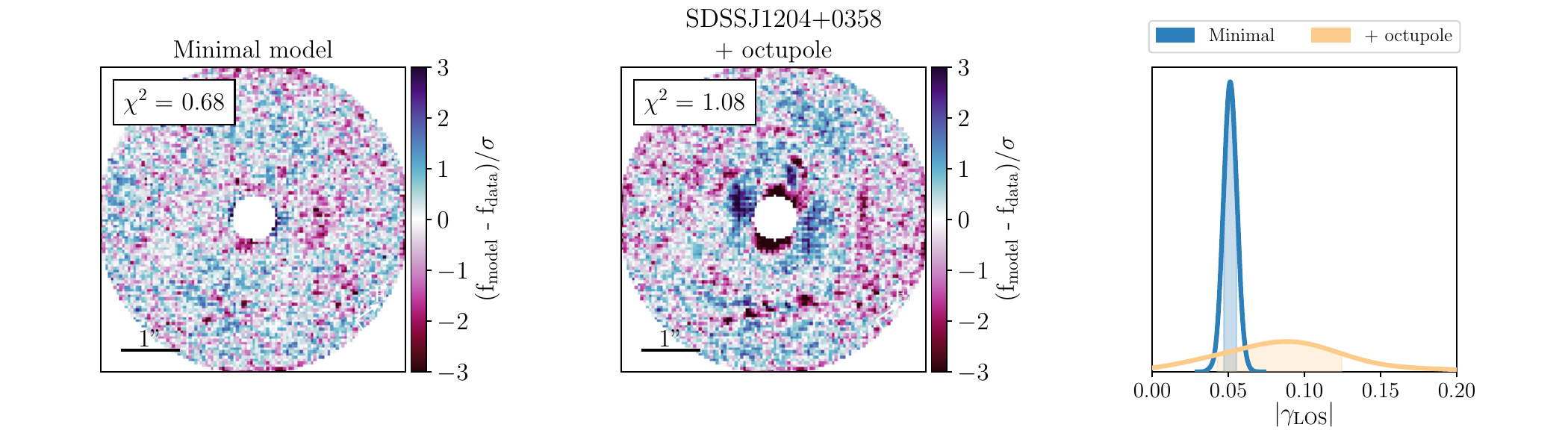}\\
	\includegraphics[width=\textwidth]{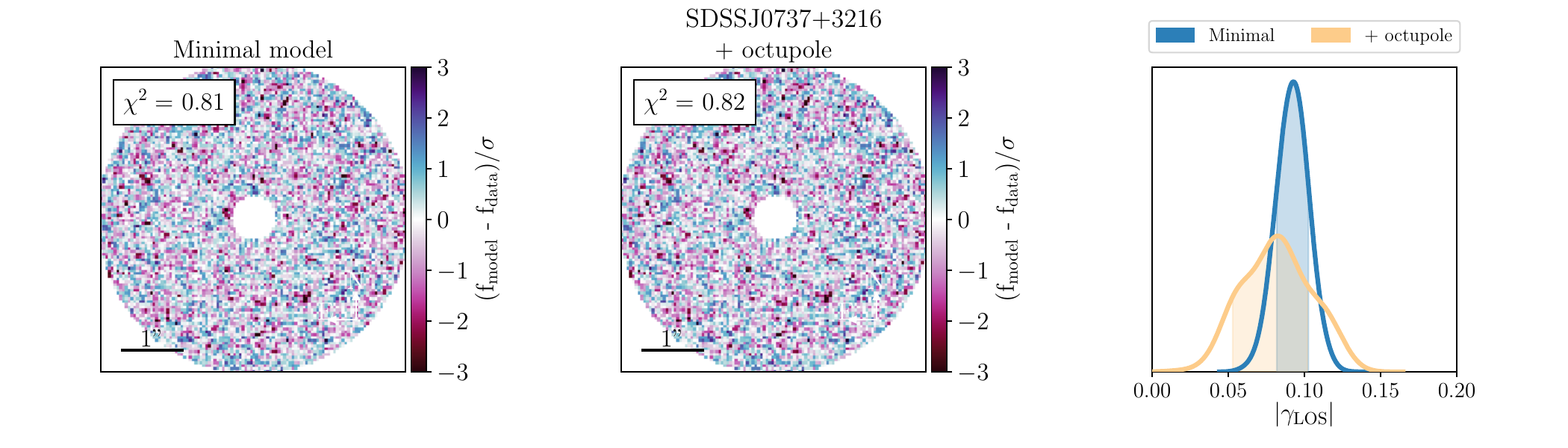}\\
	\includegraphics[width=\textwidth]{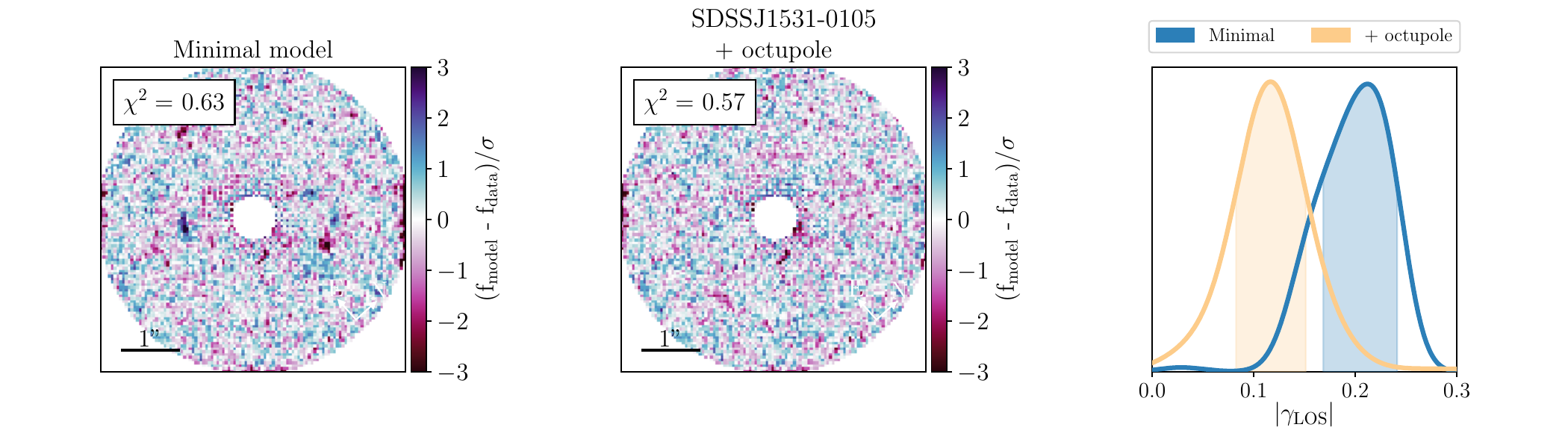}
	\caption{This figure shows the residual difference between model and data when modelling SDSSJ1204+0358, SDSSJ0737+3216 and SDSSJ1531$-$0105 using the minimal model (left panel) and with the inclusion of an octupole (middle panel). The right-hand panel shows the 1D marginalised posterior distribution of the LOS shear magnitude $|\gamma_{\rm LOS}|$ in the minimal model (blue) and with the inclusion of an octupole (yellow).}
	\label{fig:minbodi}
\end{figure*}

The final remaining three lenses, SDSSJ1204+0358, SDSSJ0737+3216 and SDSSJ1531$-$0105, show changes in tension, $\Delta\sigma$, of $-2.2$, $-3.4$ and $-4.5$ respectively. In \autoref{fig:minbodi}, we show the residual difference between data and model for the reconstructions of these lenses in the minimal model and with the octupole added to the mass model. We also show the 1D marginalised posterior distributions for the LOS shear magnitude, $|\gamma_{\rm LOS}|$, inferred in these two models (minimal model in blue, model with additional octupole in yellow).

For lens SDSSJ1204+0358, we can see two things. Firstly, with the octupole included in the model, the residuals between the data and model are significantly increased. Secondly, we can see that the apparent reduction in tension between the $|\gamma_{\rm LOS}|$ value inferred in the octupole model and the expectation from simulations for this lens is driven by a large increase in the uncertainty on the $|\gamma_{\rm LOS}|$ measurement; the error bars are over ten times larger in the octupole model than in the minimal model.

Another thing to note is that our baseline model is nested within the octupole model, meaning that, if the baseline model was the true solution to the lens equation for this system, we should expect that the octupole model would recover the baseline model via the octupole strength $a_4$ going to zero. For this lens, we find that $a_4 = -0.02^{+0.07}_{-0.04}$, which is consistent with zero at $1 \sigma$. Nevertheless, a different solution to the lens equation is clearly found, with the combination of a more concentrated source and shallower EPL slope than in the baseline model being recovered. This may point to the instability of the solutions found by either or both models.

For lens SDSSJ0737+3216, the two models give virtually identical fits to the data. Again, the reduction in tension with the simulated data is brought about by the slight loss of precision on $|\gamma_{\rm LOS}|$ when the octupole is included in the lens mass model. However, with equivalent fits and identical reduced $\chi^2$ values of $0.83$, Ockham's razor implies that, for this particular lens, we should prefer the model with the fewest parameters\footnote{The most principled way to determine the best-fitting model for each lens would be to carry out a model comparison using the Bayesian evidence, which may be obtained by using nested sampling to compute the normalising factor in Bayes' theorem, e.g. \citep{Skilling2004, Handley2015a}. An attempt at population-level lens model comparison would benefit from a larger sample of lenses than that used in this work.} i.e. the minimal model without the octupole. 

For lens SDSSJ1531$-$0105, the inclusion of the octupole serves to reduce the residuals between model and data present in the result using the minimal model, implying a better fit of the octupole model to the data. The $|\gamma_{\rm LOS}|$ value is shifted to a lower value, distinct from the result in the minimal model at just over $1\sigma$. This is the single case in our sample of lenses where the inclusion of the octupole both provides a better fit to the data than the minimal model and also reduces the LOS shear magnitude. From this we can conclude that unmodelled octupoles are not universally responsible for larger-than-expected shear magnitudes, but may be a necessary addition to a lens mass model in some specific cases.

Lastly, let us comment on the inferred octupole strength, $a_4$. In $35\%$ of the lenses studied, we recovered an octupole strength  equal to or greater than $0.02$ i.e. $2\%$, whilst a further $45\%$ have octupole strengths greater than $5\%$. This is in contrast to the findings of various observational studies of early-type galaxies, which have consistently found octupolar distortions at the $1\%$ level \citep{Hao2006, Pasquali2006, Goullaud2018}.

The inclusion of the octupole was justified in this and previous works as an aspect of mass model complexity that is well-motivated by observations. Our finding that, in the majority of the lenses studied in this work, the octupole strength, along with the shear magnitude, is larger than expected, indicates that the $m_4$ multipole alone also does not capture the full complexity of strong lens galaxy mass profiles. We emphasise that our goal is not to robustly measure the octupole itself, but rather to test whether the large recovered shear values are a symptom of its absence in other models, a question to which the answer appears to be no. We leave the exploration of other possibilities, such as the addition of an $m_1$ or $m_3$ multipole \citep{Amvrosiadis2025, Lange2025}, ellipticity gradients and isophotal twists \citep{VandeVyvere:2022gqa}, or LOS flexion \citep{Duboscq:2024asf}, to a future work.

\subsection{Neglecting post-Born corrections affects LOS shear measurements}\label{subsec:godf}

\cite{Hogg:2022ycw} demonstrated that, in mock data, removing the post-Born correction to the potential of the main deflector which is induced by foreground shear (achieved by fixing $\gamma_{\rm od} = 0$ in the parameter inference) had little effect on the inferred LOS shear magnitude. We now examine whether this finding holds true in real data; if it does, this would be useful from a parameter inference perspective, as it would reduce the dimensionality of the parameter space by two.

\begin{figure*}
	\centering
	\includegraphics[width=\textwidth]{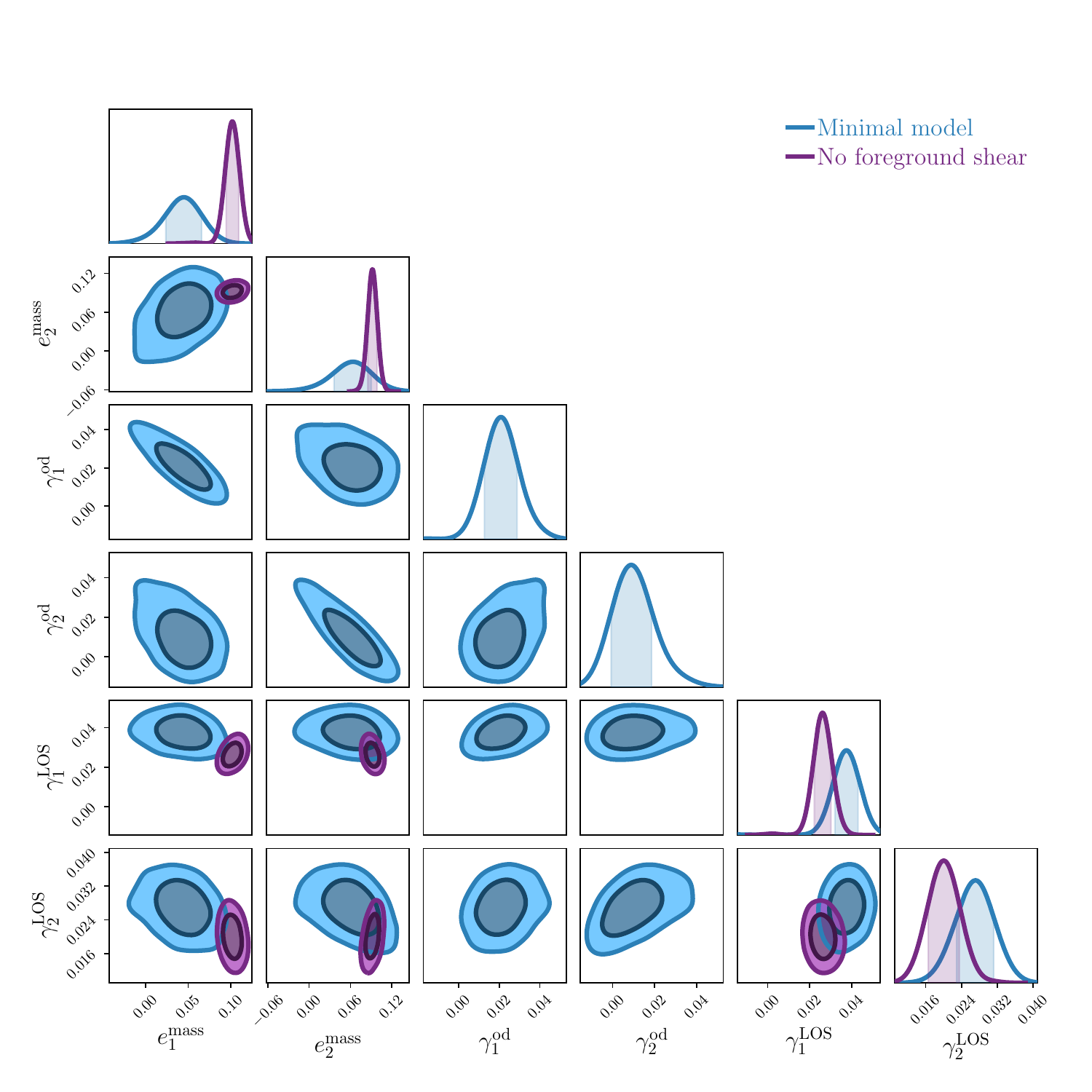}
	\caption{The one and two-dimensional joint marginalised posterior distributions for the lens mass ellipticity, foreground shear and LOS shear in lens SDSSJ1630+4520. The posteriors recovered in the minimal model are shown in blue and those recovered in the no foreground shear model in purple.}
	\label{fig:mingodf_triangle}
\end{figure*}

In \autoref{fig:mingodf_triangle}, we show the joint marginalised posterior distributions for the lens mass ellipticity components, $e_1^{\rm mass}, e_2^{\rm mass}$, the foreground shear components, $\gamma_1^{\rm od}, \gamma_2^{\rm od}$, and the LOS shear components, $\gamma_1^{\rm LOS}, \gamma_2^{\rm LOS}$ for the lens SDSSJ1630+4520. This lens was chosen as an example as the inferred LOS shear magnitude in the minimal model in this lens is close to the median for the sample, $|\gamma_{\rm LOS}| = 0.05$. We show both the posteriors recovered in the minimal model in blue and the posteriors recovered in the no foreground shear model in purple.

From this figure, we can see that when the foreground shear is excluded from the model, there are two consequences. Firstly, the inferred lens mass ellipticity is recovered with significantly higher precision than in the minimal model. Secondly, the posteriors of the LOS shear components are shifted with respect to those in the minimal model. In other words, with the enforced absence of the strong degeneracy with foreground shear, the lens mass ellipticity is recovered with artificially high precision. Globally, a slightly different best-fit lens model is therefore obtained, with a slightly different inferred LOS shear.

\begin{figure*}
	\centering
	\includegraphics[width=\textwidth]{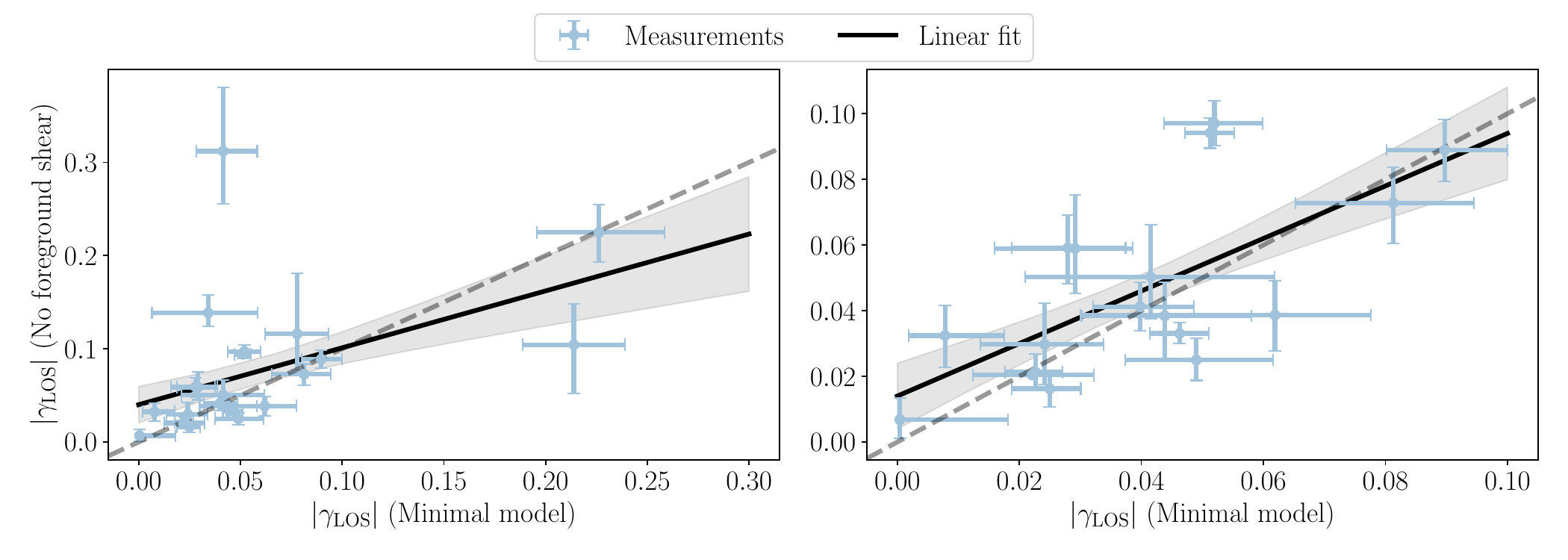}
	\caption{\textit{Left:} $|\gamma_{\rm LOS}|$ measurements in the minimal model versus the no foreground shear model. \textit{Right:} the same but restricted to show only $|\gamma_{\rm LOS}| < 0.1$. The black solid line represents a linear ordinary least squares fit to the measurements, with the $1 \sigma$ uncertainty shown by the grey shaded band. The $y=x$ line is shown by the grey dashed line.}
	\label{fig:min_vs_godf}
\end{figure*}

We can explore whether this effect is systematic across our whole lens sample. In \autoref{fig:min_vs_godf}, we show our measurements of $|\gamma_{\rm LOS}|$ in the minimal model on the $x$-axis plotted against the $|\gamma_{\rm LOS}|$ in the no foreground shear model on the $y$-axis. The left panel shows the full sample whilst the right panel restricts the view to values of the LOS shear magnitude less than 0.1, to improve the clarity of the cluster of measurements between 0 and 0.05. Furthermore, the black solid line and grey shaded bands show a linear fit\footnote{To do this, we use the ordinary least squares fitting method provided by \texttt{numpy.polyfit}. However, least squares fitting becomes inaccurate when the independent variable has significant uncertainties, as is the case here \citep{Fuller1987}. We further choose not to weight the fit using the uncertainty on the dependent variable. The displayed uncertainty on each linear fit (shaded grey region) is therefore computed from the covariance matrix returned by \texttt{numpy.polyfit} based solely on the shear values, without their errors.} and $1\sigma$ uncertainty, whilst the grey dashed line shows $y=x$.

This figure shows that for large shear values, $|\gamma_{\rm LOS}| > 0.1$, the LOS shear magnitude measured in the no foreground shear model tends to be slightly smaller than what would be measured in the minimal model, whilst at small shear values (which themselves are likely to be more trustworthy), the inverse is true: the LOS shear magnitude tends to be slightly smaller in the minimal model than when the foreground shear is neglected.

From this we may conclude that the small speed enhancement engendered by the removal of the foreground shear parameters from the inference is not worth the possible bias this will induce on the inferred LOS shear magnitude; in other words, physical accuracy should not be traded for speed.

\section{Conclusions}\label{sec:conclusions}
In this work, we modelled 23 of the SLACS strong lenses, aiming to robustly measure the LOS shear for the first time. We used a lens model which consisted of an elliptical power law profile for the main deflector, two S\'ersic profiles for the lens light, and a S\'ersic plus shapelets for the source. On top of this, we modelled the LOS shear using the minimal model of \cite{Fleury:2021tke}. We found that the distribution of our LOS shear magnitude measurements is consistent with those in the recent literature, obtained using the usual `external' or `residual' shear model.

We examined the effect of neglecting the post-Born corrections to the main deflector due to foreground shear in our model, re-fitting the same lenses with these parameters fixed to zero. We found that this leads to small changes in the inferred LOS shear magnitude compared to the minimal model, with the differences becoming more significant when the inferred shear is large.

From this, we conclude that, if shear measurements are the goal of a lens modelling programme, the minimal line-of-sight model must be used. It is too risky to neglect the foreground shear for the sake of marginally faster posterior sampling, unless the foreground is known to be insignificant. This may be true for isolated strong lens systems but is likely not the case for lenses in dense environments. We note that this conclusion may be dependent on the choice of lens mass model, and may not hold for inference performed with a different model to the elliptical power law used in this work.

We also fitted our lenses with a further degree of freedom in the lens mass model, allowing for octupolar distortions, i.e. boxyness and diskyness. We found that shear measurements are not generally reduced by the addition of this octupole, and only a single lens in our sample, SDSSJ1531$-$0105, was better fit by the octupole model than the minimal model. Nonetheless, measured shear values do in several cases show sensitivity to its inclusion.

It is also clear that the question of whether the total magnitude of the LOS shear parameter is produced by line-of-sight effects remains open. We found LOS shear measurements in good agreement with external shear measurements from the recent literature, which are known to be larger than weak lensing shear estimated around strong lenses \citep{Etherington:2023yyh}, as well as those from simulations, as discussed above. This is the key point we wish to emphasise: given the agreement between our findings and those of previous works, we conclude that the shear contribution from the lines of sight of the SLACS strong lenses studied in this work are adequately captured by the lens model.

It is the modelling of the main deflector to which attention must now be devoted, if the cosmological shear signal is to be disentangled from the contribution of the lens. However, our results also show that adding boxyness and diskyness to the lens mass model is not the panacea that might have been expected, at least for the strong lenses modelled in this work. Whether any further complexity can be added to mass models that is still physically well-motivated remains to be seen.


In conclusion, the task of uniformly modelling a very large catalogue of strong lenses and obtaining accurate shear measurements for the purposes of cosmology will clearly not be achieved with current data or modelling techniques. Lens model constraints may be improved using image data in which the lensed source light is more easily separable from the lens light; this may be seen in James Webb Space Telescope (JWST) imaging of high-redshift lensed sources  e.g. \cite{Mercier2024} -- since JWST observes in the near and mid-infrared, it has a much higher signal-to-noise ratio for redder sources compared to e.g. HST. Furthermore, more work must be done to build lens models which accurately match the physical complexity of strong lens galaxies. The promise of shear measurements from strong lensing as a novel cosmological probe motivates the pursuit of advancements in models and modelling techniques, and progress continues on each of these fronts.

\section*{Acknowledgments}

We are very grateful to Th\'eo Duboscq, Pierre Fleury and Giacomo Queirolo for useful discussions throughout the preparation of this work and for their insightful comments on the manuscript. We are also grateful to an anonymous referee whose comments led to a substantial improvement of our results. 

NBH is supported by the research environment and infrastructure of the Handley Lab at the University of Cambridge. DJ acknowledges support by the First Rand Foundation, South Africa, and the Centre National de la Recherche Scientifique of France.

\section*{Data and software availability}
The imaging data for the SLACS lenses studied in this work are publicly available. The lens modelling carried out in this work used the open-source \texttt{dolphin} package \citep{Shajib:2025bho}, with \lenstronomy \citep{Birrer:2018xgm, Birrer2021} as the modelling engine. The inputs and outputs of our modelling, including lens image cutouts, PSFs and MCMC chains, can be found at the following Zenodo repository: \url{https://zenodo.org/records/17816315}. A Jupyter notebook which reproduces the figures and tables in this work can be found at the following Github repository: \url{https://github.com/nataliehogg/los_in_slacs}.

The following colour-map packages were used in this work:  \texttt{cubehelix} \citep{Green2011},  \texttt{cmocean} \citep{Thyng2016} and \texttt{cmasher} \citep{vanderVelden2020}. NBH acknowledges the use of the AI coding agent Claude Code (Sonnet 4.5) for software development tasks. All AI-generated outputs were validated by NBH to ensure their accuracy.

\bibliographystyle{apsrev4-1}
\bibliography{los_slacs_i_oja}

\appendix\label{appendix:config}
\renewcommand{\arraystretch}{1.8}

\begin{table*}
    \centering
\begin{tabular}{llcccccc}
\hline \hline
Name & Filter & $n_{\rm max}$ & $z_{\rm lens}$ & $z_{\rm source}$ & $|\gamma_{\rm LOS}|$ & Tension [$\sigma$] & $\Delta \sigma$ \\
\hline
SDSSJ0029$-$0055 & F606W & 6 & 0.227 & 0.931 & $0.029^{+0.009}_{-0.013}$ & 1.30 & ~~0.61 \\
SDSSJ0037$-$0942 & F606W & 6 & 0.195 & 0.632 & $0.078^{+0.015}_{-0.016}$ & 3.88 & ~~0.26 \\
SDSSJ0252+0039 & F606W & 6 & 0.280 & 0.982 & $0.022^{+0.010}_{-0.010}$ & 0.85 & $-0.24$ \\
SDSSJ0330$-$0020 & F606W & 6 & 0.351 & 1.071 & $0.049^{+0.013}_{-0.012}$ & 1.99 & $-1.14$ \\
SDSSJ0728+3835 & F606W & 16 & 0.206 & 0.688 & $0.042^{+0.017}_{-0.013}$ & 2.16 & ~~1.53 \\
SDSSJ0737+3216 & F555W & 6 & 0.322 & 0.581 & $0.090^{+0.010}_{-0.010}$ & 4.86 & $-3.36$ \\
SDSSJ0903+4116 & F606W & 6 & 0.430 & 1.065 & $0.052^{+0.008}_{-0.008}$ & 1.98 & ~~3.46 \\
SDSSJ0959+0410 & F555W & 16 & 0.126 & 0.535 & $0.040^{+0.009}_{-0.008}$ & 3.34 & ~~n/a \\
SDSSJ1112+0826 & F606W & 6 & 0.273 & 0.629 & $0.226^{+0.032}_{-0.030}$ & 6.65 & $-0.57$ \\
SDSSJ1204+0358 & F606W & 6 & 0.164 & 0.631 & $0.051^{+0.004}_{-0.004}$ & 4.89 & $-2.15$ \\
SDSSJ1250+0523 & F555W & 16 & 0.232 & 0.795 & $0.025^{+0.005}_{-0.006}$ & 1.45 & $-0.15$ \\
SDSSJ1306+0600 & F606W & 6 & 0.173 & 0.472 & $0.042^{+0.020}_{-0.021}$ & 1.62 & $-0.17$ \\
SDSSJ1313+4615 & F606W & 10 & 0.144 & 0.339 & $0.024^{+0.010}_{-0.010}$ & 1.62 & ~~3.32 \\
SDSSJ1402+6321 & F555W & 6 & 0.205 & 0.481 & $0.008^{+0.010}_{-0.006}$ & 0.14 & ~~0.25 \\
SDSSJ1531$-$0105 & F606W & 6 & 0.160 & 0.744 & $0.214^{+0.025}_{-0.025}$ & 7.85 & $-4.48$ \\
SDSSJ1621+3931 & F606W & 6 & 0.245 & 0.602 & $0.028^{+0.009}_{-0.009}$ & 1.39 & ~~5.48 \\
SDSSJ1627$-$0053 & F555W & 6 & 0.208 & 0.524 & $0.023^{+0.004}_{-0.005}$ & 1.41 & ~~0.84 \\
SDSSJ1630+4520 & F555W & 6 & 0.248 & 0.793 & $0.046^{+0.005}_{-0.005}$ & 3.29 & $-0.34$ \\
SDSSJ1636+4707 & F606W & 16 & 0.228 & 0.674 & $0.062^{+0.016}_{-0.032}$ & 1.64 & $-0.81$ \\
SDSSJ2238$-$0754 & F555W & 16 & 0.137 & 0.713 & $0.034^{+0.024}_{-0.028}$ & 1.02 & $-0.42$ \\
SDSSJ2300+0022 & F555W & 16 & 0.229 & 0.464 & $0.044^{+0.014}_{-0.014}$ & 2.15 & $-0.22$ \\
SDSSJ2303+1422 & F555W & 6 & 0.155 & 0.464 & $0.000^{+0.018}_{-0.000}$ & 0.57 & ~~n/a \\
SDSSJ2343$-$0030 & F606W & 6 & 0.181 & 0.463 & $0.081^{+0.013}_{-0.016}$ & 4.17 & ~~1.73 \\
\hline
\end{tabular}
\caption{The SLACS strong lenses studied in this work. All redshifts are taken from \cite{Bolton2008}. The $|\gamma_{\rm LOS}|$ values listed are those inferred in the minimal LOS shear model. The $\Delta \sigma$ values are the change in tension with the expectation from simulations when the octupole is added to the lens model.}
\label{tab:lenses}
\end{table*}

\begin{table*}
	\centering
\begin{tabular}{llll}
\hline
\hline
Component    & Parameter        &  Prior \\
\hline
	         & $\gamma^{\rm EPL}$         & $[1.3, 2.8]$ \\
          & $e_1$            & $[-0.5, 0.5]$\\
Lens mass & $e_2$            & $[-0.5, 0.5]$\\
	         & $x$              &  $[-0.5'', 0.5'']$\\
	         & $y$              & $[-0.5'', 0.5'']$\\
\hline
             & $R_{\rm{eff}}$   & $[0.1, 5.0]$\\
	         & $e_1$            & $[-0.5, 0.5]$\\
Lens light   & $e_2$            & $[-0.5, 0.5]$\\
	         & $x$              &  $[-0.5'', 0.5'']$\\
             & $y$              & $[-0.5'', 0.5'']$\\
\hline
	         & $\beta$          & $\log_{10} [0.02'', 0.2'']$\\
	         & $R_{\rm{eff}}$   & $[0.04'', 0.5'']$\\
	         & $n_{\rm{S}}$     & $[0.5, 8.0]$\\
Source light & $e_1$            & $[-0.5, 0.5]$\\
	         & $e_2$            & $[-0.5, 0.5]$\\
	         & $x$              & $[-0.2'', 0.2'']$\\
             & $y$              & $[-0.2'', 0.2'']$\\
\hline
	         & $\gamma_1^{\rm od}$ & $[-0.2, 0.2]$\\
	         & $\gamma_2^{\rm od}$ & $[-0.2, 0.2]$\\
Shear        & $\gamma_1\h{LOS}$   & $[-0.5, 0.5]$\\
             & $\gamma_2\h{LOS}$   & $[-0.5, 0.5]$\\
	         & $\omega\e{LOS}$     & $[-0.2, 0.2]$\\
\hline
                & $a_4$   & $[-0.1, 0.1]$\\
 Octupole       & $\varphi_4$   & $[-\pi, \pi]$\\
	              & $x$      &  $[-0.5'', 0.5'']$\\
	              & $y$        & $[-0.5'', 0.5'']$\\

\hline
\end{tabular}
	\caption{Priors on the model parameters. The prior distributions are uniform between the stated limits.}
\label{tab:priors}
\end{table*}

\end{document}